\documentclass[sigconf, screen]{acmart}
\usepackage{pdfpages}
\usepackage{multirow}
\usepackage{subcaption}

\AtBeginDocument{%
  }

\copyrightyear{2026}
\acmYear{2026}
\setcopyright{cc}
\setcctype{by}
\acmConference[CHI '26]{Proceedings of the 2026 CHI Conference on Human Factors in Computing Systems}{April 13--17, 2026}{Barcelona, Spain}
\acmBooktitle{Proceedings of the 2026 CHI Conference on Human Factors in Computing Systems (CHI '26), April 13--17, 2026, Barcelona, Spain}
\acmPrice{}
\acmDOI{10.1145/3772318.3791877}
\acmISBN{979-8-4007-2278-3/2026/04}

\begin{document}

\title[Designing Proactive Functionality for In-Vehicle Conversational Assistants]{ProVoice: Designing Proactive Functionality for In-Vehicle Conversational Assistants using Multi-Objective Bayesian Optimization to Enhance Driver Experience}

\author{Josh Susak}
\orcid{0009-0007-3522-1005}
\email{joshua.susak.24@ucl.ac.uk}
\affiliation{%
  \institution{UCL Interaction Centre}
  \city{London}
  \country{United Kingdom}
}

\author{Yifu Liu}
\email{yifu.liu.22@ucl.ac.uk}
\orcid{0009-0009-5823-1411}
\affiliation{%
  \institution{UCL Interaction Centre}
  \city{London}
  \country{United Kingdom}
}

\author{Pascal Jansen}
\email{pascal.jansen@uni-ulm.de}
\orcid{0000-0002-9335-5462}
\affiliation{%
  \institution{Institute of Media Informatics, Ulm University}
  \city{Ulm}
  \country{Germany}
}
\affiliation{%
  \institution{UCL Interaction Centre}
  \city{London}
  \country{United Kingdom}
}

\author{Mark Colley}
\orcid{0000-0001-5207-5029}
\email{m.colley@ucl.ac.uk}
\affiliation{%
  \institution{UCL Interaction Centre}
  \city{London}
  \country{United Kingdom}
}

\renewcommand{\shortauthors}{Susak et al.}

\begin{abstract}
The next step for In-vehicle Conversational Assistants (IVCAs) will be their capability to initiate and automate proactive system interactions throughout journeys. However, diverse drivers make it challenging to design voice interventions tailored towards individual on-road expectations. This paper evaluates the effectiveness of Human-in-the-Loop (HITL) Multi-Objective Bayesian Optimization (MOBO) in design by implementing ProVoice: a Virtual Reality (VR) driving simulator integrating MOBO to investigate the effects of IVCA design variants on perceived mental demand, predictability, and usefulness. By reporting the Pareto Front from a within-subjects VR study (N=19), this paper proposes optimal design trade-offs. Follow-up analysis demonstrates MOBO’s success in discovering effective intervention strategies, with reduced participant mental demand, alongside enhanced predictability and usefulness while engaging with the proactive IVCA. Implications for computational techniques in future research on proactive intervention strategies are discussed. ProVoice can extend to include alternative design parameters and driving scenarios, encouraging intervention design on a broad scale.

\end{abstract}



\maketitle

\section{Introduction}
In-vehicle Conversational Assistants (IVCAs) play a fundamental role in facilitating communication between driver and vehicle, promoting ubiquitous access to system functionalities including navigation re-routing, handling calls or messages, and interior climate adjustments~\cite{weng2016conversational}. As vehicles become intelligent and software-driven, Original Equipment Manufacturers (OEMs) shift attention towards providing tools to re-imagine the Driver-Vehicle Interaction (DVI) experience. Already, IVCAs, including BMW’s Personal Intelligent Assistant~\cite{bmwPersonalAssistant} and Mercedes-Benz’s MBUX~\cite{mbux}, integrate Generative Artificial Intelligence agentic capabilities, allowing for two-way conversation while holding promise of obtaining “comprehensive and personalised information within seconds”~\cite{mercedesSearch}.

However, many commercial IVCAs operate reactively and require some activation command to converse with the driver (see MBUX's "Hey, Mercedes"~\cite{mbux} or Škoda's "Okay Laura"~\cite{skodaautokodaLaura}). The next milestone towards achieving seamless communication between driver and IVCA will be their potential to identify, initiate, and automate proactively, at opportune points throughout driving journeys. Such services aim to assist with vehicle-centric tasks and responding to in-vehicle or on-road events with the hope of advancing safe, immersive future driving experiences.

Drivers' on-road preferences are diverse~\cite{yarlagadda2022heterogeneity, tawfik2012human, ellison2013examining, ossen2007driver, gkouskos2014drivers}, while the focus demanded from driving manoeuvres can limit availability to communicate with proactive features~\cite{semmens2019now}. Designing IVCA variants can be challenging, requiring careful trade-offs in multiple opaque and subjective design objectives such as system trust and usefulness~\cite{ghazizadeh2012extending}. User-led adaptive approaches to personalisation~\cite{normark2015design, hui2015enhancing} can restrict design to a discrete set of variants identified by the end-user, with potential to overlook variations which may better match their preferences. While proactive system behaviour is characterised by the timing and relevance of the intervention, a critical question arises of how automotive manufacturers should approach the challenge of designing proactive intervention, in line with individual end-user needs and desired autonomy. By tailoring the proactive IVCA towards driver preferences, the DVI experience can be enhanced, leading to increased engagement and use of in-vehicle services.

This work proposes ProVoice, a 3D Virtual Reality (VR) driving simulator that leverages Human-in-the-Loop (HITL) Multi-Objective Bayesian Optimization (MOBO) to investigate its effectiveness for modelling proactive IVCA intervention. Design parameters focused on were: Level of Autonomy (LoA), symbol transparency, alert volume, and interior glow, to explore alternative modalities of assistant intervention, including visual and auditory communication. These parameters were optimized for three design objectives: mental demand, predictability, and usefulness. The assistant is modelled across five LoA: User performs task without system support, User performs task with system support, System performs task with user approval, System performs task with pending user veto, System performs task.

Proactive personality, or disposition, is defined as "taking initiative to influence one’s environment and bring about meaningful change"~\cite{bateman1993proactive}. In a within-subjects VR driving simulator study exploring the impact of training compared to fixing LoA based on proactive dispositional traits, N=19 participants explored ProVoice across a navigation re-routing scenario. The research questions of the study are as follows:
\begin{enumerate}
    \item \textit{How does proactive intervention design produced by Human-in-the-Loop Multi-Objective Bayesian Optimization impact driver perception of predictability, mental demand, and usefulness towards the IVCA across iterations?}
    \item \textit{Does predictability, mental demand, and usefulness change when Level of Autonomy is fixed based on proactive disposition, compared to training as a design parameter?}
\end{enumerate}

Throughout, the IVCA intervenes and adapts to the situation, offering support based on destination proximity and assigned LoA. By discovering Pareto-optimal values across two experiment conditions (Trained LoA, Fixed LoA), this paper visualises the Pareto Front. The study found reports of reduced mental demand, alongside enhanced predictability and usefulness while interacting with the IVCA and engaging with HITL MOBO across both conditions. An open-source Unity asset for modelling proactive behaviour using MOBO is released publicly, as used by ProVoice. This asset provides a baseline to integrate novel proactive IVCA intervention by including new design parameters, objectives, and driving scenarios.


\medskip

\noindent\textbf{Contribution statement}~\cite{wobbrock2016research}:
\begin{enumerate}
    \item \textbf{Artifact or System.} ProVoice: A driving simulator using HITL MOBO to explore proactive IVCA design across four design parameters and five LoA. We open-source the Unity asset for modelling proactive functionality using HITL MOBO, with a tutorial on how to extend with additional parameters and objectives.
    \item \textbf{Empirical study that tells us about how people use a system.} Empirical insights from a within-subjects study (N=19), evaluating participant feedback about the proactive intervention presented to them based on mental demand, predictability, and perceived usefulness.
\end{enumerate}

\section{Related Work}
Proactive behaviour is defined as “taking action by causing change and not only reacting to change when it happens\footnote{\url{https://dictionary.cambridge.org/dictionary/english/proactive}; Accessed: 11.09.2025}. Proactive behaviour in Conversational Assistants (CAs) is modelled by two characteristics: anticipating user intentions without waiting for a command (assumption) and achieving a given goal independently, without human intervention (autonomy)~\cite{grant2008dynamics}. While manufacturers explore proactive behaviour for non-critical driving tasks, seen in Google’s predictive route adjustment~\cite{googleRouteOptimization} or BMW’s Intelligent Personal Assistant~\cite{bmwPersonalAssistant}, many commercial assistants converse reactively. Drivers must initialise conversation using activation commands, rather than the assistant acting on their behalf.

\subsection{Proactivity in Voice Assistants}

While reactive communication reduces the potential of being surprised, it restricts additional features offered to drivers~\cite{semmens2019now}, limiting the IVCA's ability to understand broader contextual cues beyond initiated conversation. Accounting for this limitation, proactive IVCAs are considered to allow ubiquitous access to future in-vehicle services~\cite{lugano2017virtual, mathis2023exploring}. Proactivity in automotive settings is a “personification of the car’s intelligence”, holding the promise of enhancing convenience by offering relevant suggestions~\cite{mathis2024approach} throughout journeys.

Commercial IVCAs largely communicate in voice. Prior research positions voice-based input as a safe in-vehicle interaction method~\cite{sodnik2008user, mehler2016multi, ranney2005effects}. For instance, \citet{sodnik2008user} discovered drivers exhibited fewer driving distractions and minimised attentional shifts from the road while using speech-based methods, while completing secondary tasks like writing text messages or calling.
Aligning with these results, \citet{mahajan2021exploring} found active listening and conversation to combat fatigue, with participants reporting significantly higher cognitive workload, alongside lower self-reported sleepiness on the Karolinska Sleepiness Scale (KSS)~\cite{aakerstedt1990subjective}, implying higher levels of alertness following voice interaction.

Despite these opportunities, there are some crucial limitations of voice-based proactivity. In a study concerning example cases for proactive smart speakers, over 25\% of participants held regard for privacy concerns, due to the assistant needing to analyse the environment, leading to doubts about how data is handled~\cite{reicherts2021may}. In a follow-up experiment exploring acceptable circumstances for proactive intervention, participants similarly disclosed worries with privacy and mistrust, demanding further transparency in data collection~\cite{zargham2022understanding}. Such findings underscore the key dilemma concerning proactive intervention. Driver activities need to be monitored to determine opportune points of interaction, posing risks to personal privacy. However, it may be vital for an IVCA to collect as much necessary information relevant to its functioning.


\subsection{Driver-Vehicle Communication}

One challenge is identifying if, when, and how an IVCA should suggest, or initiate, proactive intervention for the driver. Timing challenges arise from the dynamic nature of the driving environment, where on-road situations such as high-density traffic can shift the workload of interacting drivers~\cite{bongiorno2017driver}. 

Through adopting a wizard-of-oz approach, Semmens et al. produced a dataset of self-annotated driver responses with 2,734 instances of proactive intervention. They uncovered higher stress and reduced communication availability before complex driving manoeuvrers (approaching a junction or roundabout)~\cite{semmens2019now}, compared to better-timed intervention points which included cruising straight in lighter traffic. Unsuitable intervention points were often caused by unexpected environmental factors beyond the driver’s control, such as uncommon situations (“Someone almost just backed into me") or vulnerable road users like cyclists~\cite{semmens2019now}.

The degree of proactive intervention - the extent to which the proactive assistant takes over - is usually presented via the LoA. Across the literature, there is no clear consensus on the number of defined stages. For example, Kraus et al. expand on ten different LoA ranging from 'no assistance' to 'complete autonomy'~\cite{kraus2021role}, whereas Du et al. formulate five LoA from 'no assumptions' to 'strong assumptions'~\cite{du2024towards}. 



This paper follows a 5-stage model with the following stages:
User performs task without system support (Level 0), User performs task with system support (Level 1), System performs task with user approval (Level 2), System performs task with pending user veto (Level 3), System performs task (Level 4).

\subsection{User Reaction towards Proactive Intervention}

Prior literature has explored how information from a proactive IVCA is conveyed to the driver, with a focus on the driver’s subjective feelings.
The relevance of the intervention to the current driving context is demonstrated to affect the driver’s willingness to interact with the IVCA. A wizard-of-oz study by Schmidt et al. reveals that, on average, 82.5\% of participating drivers were likely to react faster to approve the proactive intervention when directly related to their driving journey (re-routing, finding a parking space or stopping for re-fueling) in comparison to indirectly related non-critical tasks (scheduling a work appointment or taking a break due to tiredness)~\cite{schmidt2020users}.

Additionally, the design of proactive communication plays an important role in drivers' engagement. A crowdsourcing study by Meck highlights the need for voice-based proactive IVCAs to be authentic and human-like, with participants finding linguistically simple, polite suggestions to be most easily understandable amidst driving~\cite{meck2024may}. Similarly, Braun et al.’s real-world study which adjusted personality according to McCrae's Big Five model~\cite{mccrae1992introduction} suggests the value of tailoring assistant output to driver disposition, demonstrating higher levels of perceived trust and reliability when assistance matched participant personality~\cite{braun2019your}.

\subsection{Computational Methods for Personalised In-Vehicle User Interface Design}


In-vehicle User Interface (UI) design involves adjusting multiple design parameters to maximise or minimise design objectives like perceived usefulness or product trust~\cite{ghazizadeh2012extending}. These can be represented as objective functions, whose values are derived from a combination of parameter values. The challenge of discovering value combinations can be reframed as an optimization problem, given their opaque relationship to the objective functions~\cite{colley2025improving}.

Prior research has explored user-led adaptive approaches towards in-vehicle personalisation. For instance, Normark allowed drivers to customise icon placement and colour within the vehicle’s infotainment and Heads-Up Displays~\cite{normark2015design}. While participants disclosed a higher sense of ownership~\cite{normark2015design}, a user-controlled approach to parameter configuration can limit designers and end-users to a discrete combination set. This could lead to overlooking unexplored variants, which may achieve increased performance across objectives. Further complications arise when balancing simultaneous objectives, where modifying one may lead to a trade-off in another.

Steering away from discrete configuration, computational techniques such as HITL MOBO are explored in Driver-Vehicle Interaction, with notable applications including in-vehicle information visualization and external communication of vehicle intentions for pedestrians and onlookers~\cite{colley2025improving, jansen2025opticarvis}. By approaching UI problems under a HITL approach, MOBO holds promise for systematically exploring continuous design spaces through refining a model of the design space by directly involving end-users and optimizing based on iterative feedback, regardless of their prior design expertise or technical background. The goal of MOBO is to find the combination of design parameter values on the Pareto front. Each combination on this front is referred to as Pareto-optimal. These represent the most effective trade-offs when balancing the design objectives~\cite{jansen2025opticarvis, ngatchou2005pareto}. Jansen, Colley et al.'s OptiCarVis study exploring visualization design for automated vehicle functionality demonstrates the success of HITL feedback as a user-centred design approach in finding Pareto-optimal parameter configurations for situation detection, prediction, and trajectory planning while maximising participants' perceived sense of system safety, trust, predictability, and acceptance in a relatively low number of MOBO iterations~\cite{jansen2025opticarvis}.

Thus, HITL MOBO can combine user-centred design principles like continuous end-user feedback with progressive exploration of the design space, enabling expert designers and non-expert end-users to explore the effects of parameter configurations on high-level, opaque objectives.

\subsection{Study Gap}

A manual approach to in-vehicle personalisation becomes inefficient when considering simultaneous design objectives, with potential to overlook optimal design variants. A HITL approach can directly engage end-users and encourage in-depth exploration through the design space to uncover IVCA variants tailored towards their preferences, regardless of prior design or technological expertise. Thus, a rich exploration towards understanding optimal configurations using intelligent machine-learning techniques could inform future work into proactive intervention design.

In addition, there does not yet exist a free and open-source tool that allows for the integration of proactive system design across different scenarios, to the best of our knowledge. Such a tool could allow for future implementation of proactive functionality within a controlled environment.

The present study bridges this gap by exploring driver personalisation of a proactive IVCA, through leveraging MOBO as an interaction design technique. We present an implementation and evaluation of a custom VR driving simulator for modelling proactive IVCA design across five LoA, using an example navigation re-routing scenario. The research also considers multiple design objectives, including mental demand, predictability, and usefulness, exploring a design space with continuous parameter values.

\section{Method}

\subsection{Driving Simulator Prototype}

This project developed ProVoice, a VR driving simulator in Unity~\cite{unityUnityRealTime}. Unity offers several advantages for development, including an ecosystem of pre-built assets and plug-ins tailored for desktop and VR development. This section will discuss the design and implementation of ProVoice.

\begin{figure*}[ht!]
  \centering
  \includegraphics[width=\textwidth]{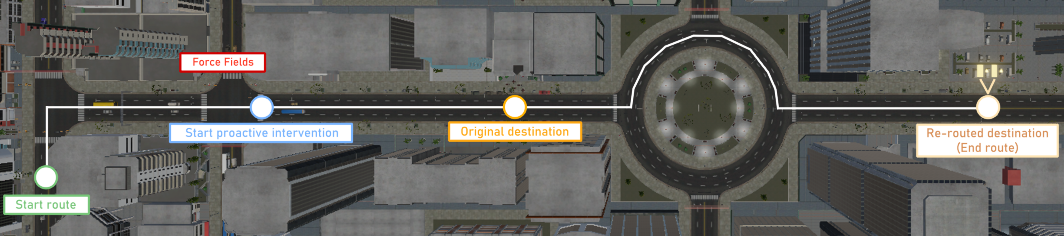}
  \caption{Route overview. White line indicates the full route for one iteration. Golden squares represent potential parking locations.}
   \Description{ProVoice Route Overview, outlining the whole route taken by drivers for one iteration.}
  \label{fig:route}
\end{figure*}

\begin{figure}[ht!]
  \centering
  \includegraphics[width=0.9\linewidth]{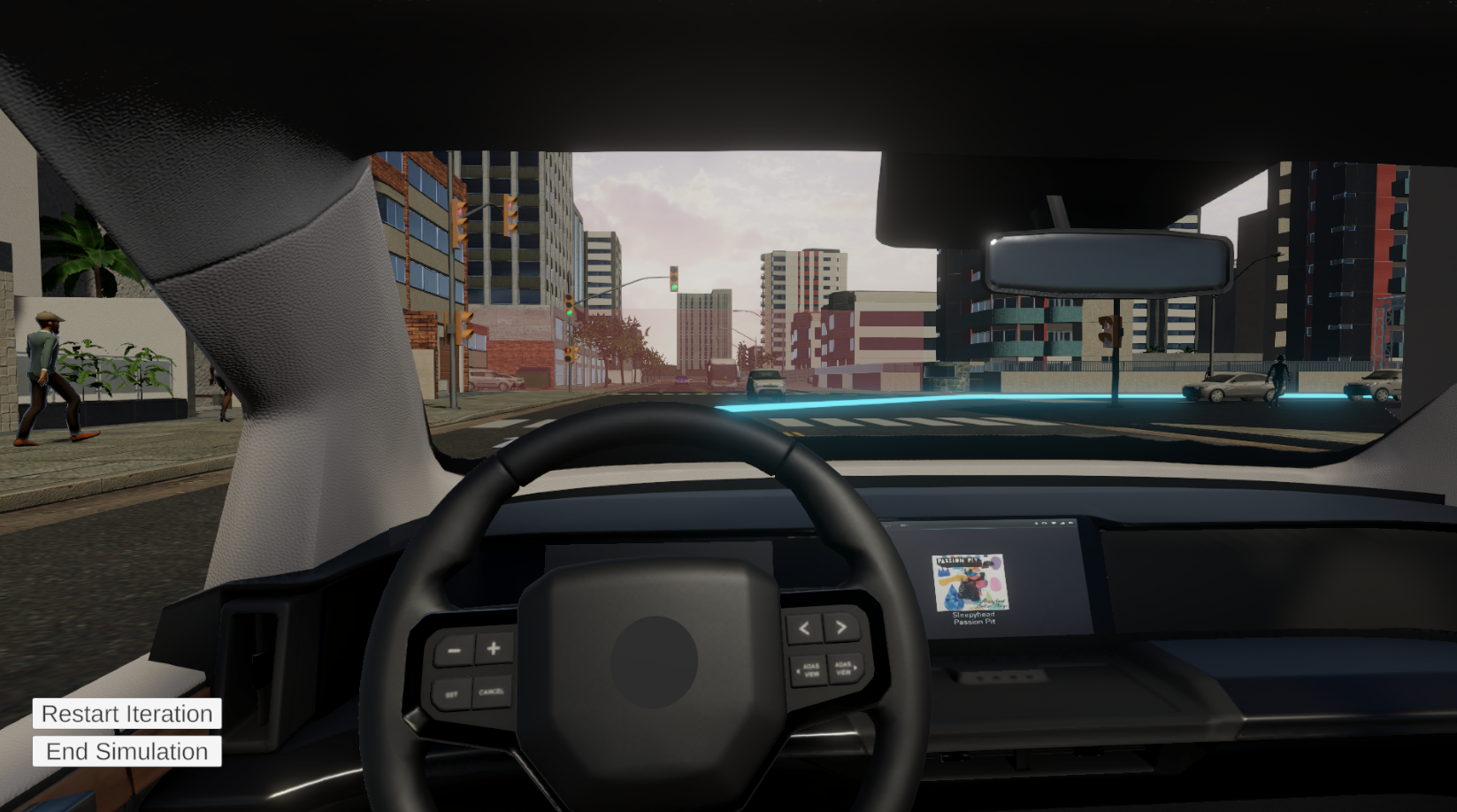}
  \caption{Interior vehicle view from the driver cockpit. The blue line represents a support path, defined as a marker to keep drivers on-course to the route boundaries.}
   \Description{Interior vehicle view from the driver's cockpit. The blue line represents a support path, keeping drivers on-course to the route boundaries.}
\end{figure}

\begin{figure}[ht!]
  \centering
  \includegraphics[width=0.9\linewidth]{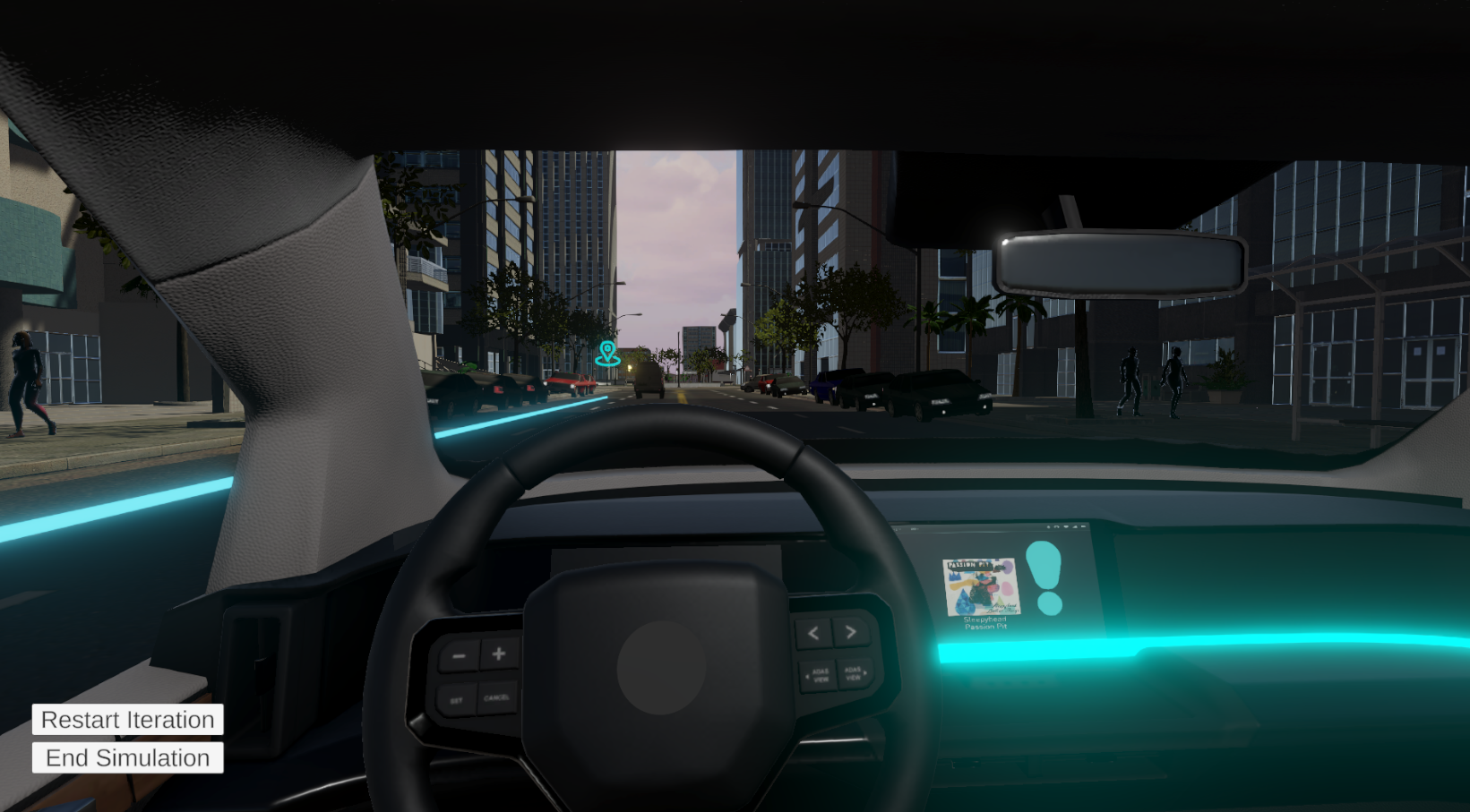}
  \caption{Interior view of the vehicle upon proactive intervention. Each iteration modifies some design parameter for the IVCA.}
    \Description{Interior view of the vehicle after proactive intervention. Upon activation, the LED strip glows cyan.}
\end{figure}

\begin{figure}[ht!]
  \centering
  \includegraphics[width=0.9\linewidth]{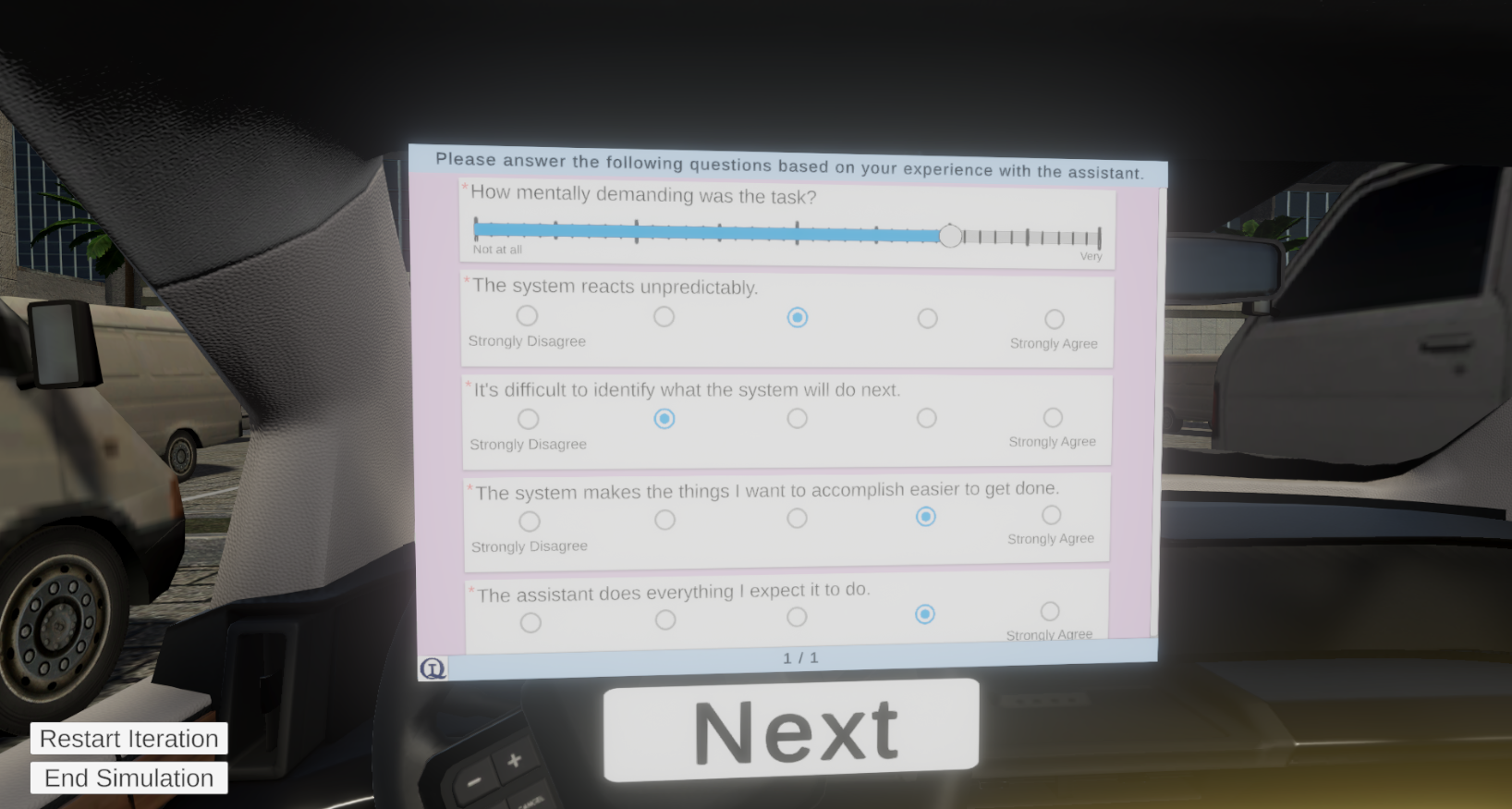}
  \caption{Interior questionnaire, presented after locating a final parking destination. Drivers are asked to rate the IVCA variant on mental demand, predictability, and usefulness.}
    \Description{Interior questionnaire, presented after locating a final parking destination. Drivers are asked to rate the IVCA variant in light of mental demand, predictability, and usefulness.}
\end{figure}

\subsubsection{Scene Design}

Prior in-situ studies draw attention to driving manoeuvre complexity as a key factor affecting proactive communication acceptance~\cite{semmens2019now, wu2021learning}. For instance, Semmen et al.’s driving study over a 16-mile journey illustrates the demands of different manoeuvres to impact driver availability, ranging from right turns to freeway merging~\cite{semmens2019now}. Therefore, the driving environment resembles an urban route composed of varying on-road conditions, including roundabouts, straight roads, intersections, and pedestrian crossings.

A view of the driving scene and route map is shown in \autoref{fig:route}.

\subsubsection{Simulator Controls}

The controls are realised through a steering wheel and pedals located at the feet. To mimic the feeling of sitting in a real vehicle cockpit, controls are mapped to semantically relevant locations familiar to a driver. This includes the gas pedal for acceleration, brake pedal for slowing down and steering to control front-wheel movement.


While posing risks to external validity due to over-simplifying movement, in cases where the researcher may not have access to driving apparatus, the simulator operates using a standard gaming controller. In all experiments, participants controlled ProVoice using Logitech MOMO driving apparatus rather than gamepad or keyboard controls.

\subsubsection{Driving Task}

The driver must follow a given route, relayed through a GPS navigation system. Based on Müller, Colley et al.'s exploration into in-vehicle Augmented Reality (AR) visualizations, the navigation line was made thin and placed at a low height on the driver's side of the road to reduce visual obstruction~\cite{muller2022ar4cad}. To finish one iteration, they must reach an allocated parking destination, indicated by a golden collider.
During the journey, the proactive IVCA intervenes, providing a suggestion or automating an action according to the assigned LoA. The driver can interact with the IVCA using voice commands, during a short grace period. Drivers were asked to answer a built-in questionnaire after every iteration to inform optimization. Their feedback is used to adjust design parameters. The scenario then repeats under a different IVCA variation.


\subsubsection{Proactive Scenario Design}

As presented by Schmidt et al. in their exploration of driving-related proactive use cases, 20 participants viewed parking re-routing as bringing most additional consumer value to the assistant~\cite{schmidt2019exploration}. These thoughts were validated in a subsequent on-road wizard-of-oz study, where the re-routing intervention was accepted in 95\% of all triggers~\cite{schmidt2020users}. As the scenario with the highest potential real-world driver acceptance, this paper explores a navigation re-routing scenario. 
The full scenario is shown in Appendix~\ref{appendix:intervention}.



\subsection{Multi-Objective Bayesian Optimization Setup}

\begin{table*}
\small
\centering
\caption{The four design parameters used for proactive IVCA design and their range.}
\begin{tabular}{llll}
\toprule
\textbf{Design Parameter} & \textbf{Description} & \textbf{Ref.} & \textbf{Range} \\
\midrule
$p_1$: Interior Lighting Glow & Glow intensity of the interior dashboard lighting. &~\cite{shah2020vehicle} & [0, 1] \\
$p_2$: Auditory Alert Volume & Volume of the auditory alert. &~\cite{wiese2004auditory} & [0, 1] \\
$p_3$: Symbol Transparency & Transparency of the infotainment intervention symbol. &~\cite{shah2020vehicle} & [0, 1] \\
$p_4$: Level of Autonomy, LoA & LoA used to initiate, or automate, a proactive event. & & [0, 1]: 5 steps \\
\bottomrule
\label{table:parameters}
\end{tabular}
\end{table*}

\subsubsection{Design Parameters}

Design parameters for modelling the proactive IVCA were derived from~\citet{shah2020vehicle, wiese2004auditory}, as shown in \autoref{table:parameters}. Each parameter is modelled within a continuous range $v$ $\in$ [0,1], to denote visibility. 0 and 1 are included to allow for unwanted elements to be completely added or removed. Red, blue, and green were considered as parameters to personalise visual intervention colour but were avoided due to potential clashing with in-vehicle functionality (for example, red to indicate immediate error~\cite{theaaDashboardWarning}), which could lead to unrelated meaning. All visual intervention displays a bright cyan hue. As a colour outside the red-yellow-green spectrum, cyan is used to direct attention towards informational cues~\cite{locken2016enlightening,muller2022ar4cad}. In addition, Colley, Jansen et al.'s study on Bayesian Optimization for external communication of automated vehicles found hues close to cyan to be consistent across many Pareto-optimal designs~\cite{colley2025improving}.

\subsubsection{Design Objectives \& Human-in-the-Loop Feedback}

Taken from the Automation Acceptance Model~\cite{ghazizadeh2012extending}, the simulator considers two design objectives to maximise: predictability and perceived usefulness. Mental demand was the sole objective to minimise. 
\textbf{Mental demand} is assessed using the mental workload subscale from Hart and Staveland’s NASA Task Load Index (TLX)~\cite{hart1988development} (“How mentally demanding was the task?”: 0 = Not demanding at all, 20 = very demanding).
\textbf{Predictability} is measured with two inverse likert scale questions, using predictability subscales from Körber’s trust in automation survey~\cite{korber2018theoretical} (“The system reacts unpredictably.”: 1 = Strongly disagree, 5 = Strongly agree) (“It's difficult to identify what the system will do next.”: 1 = Strongly disagree, 5 = Strongly agree). All scores for predictability were reversed prior to data analysis.
\textbf{Usefulness} is recorded with two likert scale questions adapted from Lund's Usefulness, Satisfaction and Ease of use (USE) questionnaire~\cite{lund2001measuring} (“The system makes the things I want to accomplish easier to get done.”: 1 = Strongly disagree, 5 = Strongly agree) (“The assistant does everything I expect it to do.”: 1 = Strongly disagree, 5 = Strongly agree).

\subsubsection{Hyperparameter Setup for Bayesian Optimization}
This study integrates the Unity Bayesian Optimization plugin by Jansen, Colley et al.\cite{githubGitHubPascalJansenBayesianOptimizationforUnity}, with BoTorch\cite{balandat2020botorch} and applies q-Noisy Expected Hypervolume Improvement (qNEHVI) as the acquisition function. Following established HITL MOBO practice in in-vehicle UI design~\cite{jansen2025opticarvis,chan2022investigating}, a batch size of $q{=}1$ ensures that one design is evaluated per iteration. A seed of 3 supports reproducibility, and the acquisition is approximated using 1024 restart candidates and 512 Monte Carlo samples~\cite{jansen2025opticarvis,chan2022investigating}.

The MOBO setup follows prior work demonstrating that Gaussian–process–based optimization can identify user-tailored in-vehicle UI designs within small iteration budgets, even under subjective noise and conflicting design objectives~\cite{jansen2025opticarvis}. Alternative computational techniques were examined, but were not suitable for this problem setting. Supervised learning approaches typically require large, labeled datasets that map design parameters to subjective ratings~\cite{fiebrink2011real,alonso2015challenges}, which are impractical to obtain for proactive IVCA design due to the strong dependence on situational driving context and intra-individual variability. Reinforcement learning typically requires predefined reward structures and long interaction sequences to learn effectively~\cite{arzate2020survey,jokinen2021touchscreen}, which conflicts with the limited, fatigue-sensitive evaluation cycles feasible in VR driving studies. Evolutionary and genetic algorithms offer broad search capabilities but often exhibit unstable performance with noisy human-generated feedback, requiring substantially larger sample budgets than are realistic in HITL experimentation~\cite{katoch2021review}.

In contrast, MOBO is explicitly designed for optimization problems with sparse, noisy, and expensive observations~\cite{shahriari2015taking}. A multi-output Gaussian process surrogate is used to model the three design objectives jointly. The Gaussian process treats each objective as a smooth but unknown function over the design parameters, providing both predictions and uncertainty estimates that allow the optimizer to balance \textit{exploration} (i.e., testing designs in uncertain regions of the space) and \textit{exploitation} (i.e., refining designs that currently appear promising) efficiently.

Design selection in each HITL iteration is governed by BoTorch’s qNEHVI acquisition function, which is specifically designed for multi-objective problems with heterogeneous observation noise and performs reliably in small-iteration HITL settings (e.g., 15 iterations~\cite{jansen2025opticarvis}). We expected noise for subjective ratings of mental demand, predictability, and usefulness, likely due to intra-individual variability, route context, habituation, and repeated questioning across iterations. The MOBO setup addresses this through a multi-output Gaussian-process surrogate that explicitly models noisy observations and propagates this uncertainty into the qNEHVI acquisition step. This combination enables the optimizer to distinguish genuine preference structure from measurement fluctuations, avoids scalarizing the objectives, and supports efficient, noise-aware exploration of the IVCA design space within the practical constraints of VR-based driving studies.

\subsection{Driving Simulator Study}

The driving simulator study evaluates the impact of HITL MOBO for modelling optimal IVCA designs, particularly considering variations when LoA is trained using MOBO compared to fixing according to proactive disposition.

\begin{figure}[ht!]
  \centering
  \includegraphics[width=.45\textwidth]{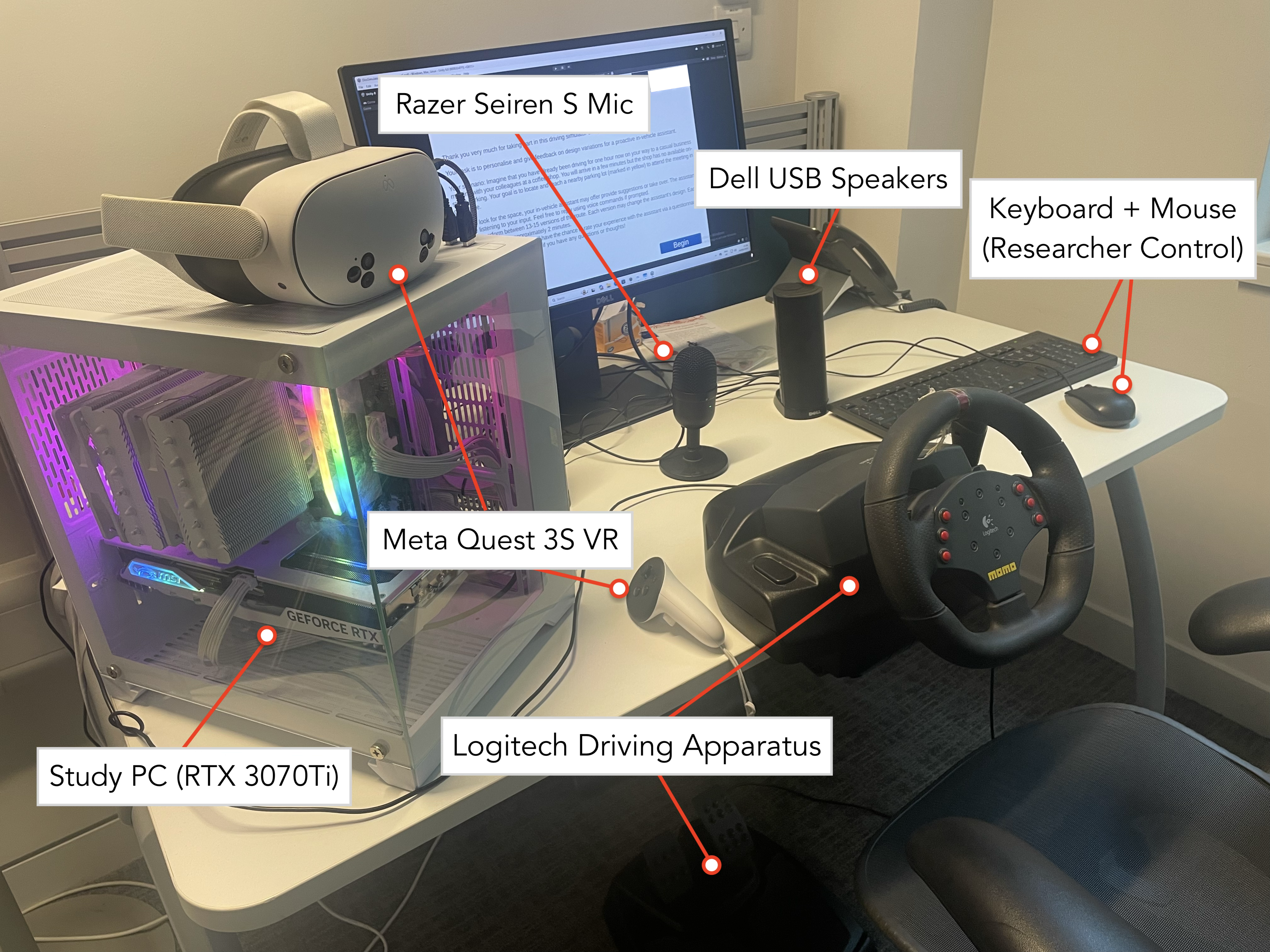}
  \caption{Exemplar experiment setup. Participants controlled ProVoice with third-party driving equipment and VR headset. The simulator was displayed on a desktop monitor for direct researcher observation.}
   \Description{Exemplar experiment setup. Participants controlled ProVoice with third-party driving equipment and VR headset. The simulator was displayed on a desktop monitor for direct researcher observation.}
  \label{fig:setup}
\end{figure}

\subsubsection{Participants}

Given that on-road experience and driving heterogeneity are crucial factors affecting driver preferences towards potential proactive intervention designs, the study recruited people with baseline knowledge of operating a vehicle under a valid provisional or full driving license, obtained within the UK or internationally. Participants were also required to meet key eligibility criteria, including: Being at least 18, being able to effectively read, understand, write, and speak English; being able to provide verbal and written consent; and not considering themselves to be a vulnerable adult. 

19 participants in total (10 male, 9 female, Mean Age = 31.4, SD = 11.3, Range = [19-54]) (Education: 6 Bachelor's degrees, 4 Master's degrees, 5 Some college/university (no degree), 4 High school or equivalent) were recruited through SONA systems and personal advertisement. All participants met the screening criteria of holding a full or provisional driving license. Average score for proactive disposition was 87.9 (SD = 10.5, Range = [69-102]). The 7 participants holding a provisional license self-disclosed their on-road driving experience upon signing up to the study, as stated on the recruitment sheet (4 a few times a year, 3 a few times a month). 

\subsubsection{Study Design}

To compare the effects of fixing versus optimizing LoA while interacting with HITL MOBO (RQ2), a within-subjects study design was used. Participants experienced two experiment blocks in a counterbalanced order, with a short break to reduce order effects and confounding variables. The two main conditions are: 

\begin{enumerate}
    \item \textbf{Trained LoA}, as a fourth design parameter.
    \item \textbf{Fixed LoA}, based on participant disposition to proactive behaviour.
\end{enumerate}

The independent variable was the type of LoA used during proactive intervention (trained or fixed). The dependent variables were the participant questionnaire scores throughout iterations, used to optimize the assistant and record subjective measures towards the design objectives.

\subsubsection{Materials}

The experiment took place in a quiet and well-lit room. The room was air-conditioned to combat potential feelings of nausea arising from extended VR use. \autoref{fig:setup} demonstrates the full annotated study setup. Participants controlled the simulator using Logitech MOMO driving controls, with force-feedback and 170 degrees of rotation. The steering wheel was clamped to the centre of the table. The pedals were placed in a comfortable position at the participant's feet.
A Meta Quest 3S VR headset connects to the desktop computer through a USB-C cable to ensure a stable streaming connection. A VR controller was placed to the side of the participant to pick up between iterations. The driving simulator is powered by Unity 6000.0.47f1 and maximised on the desktop monitor. This provided a dual-screen view for the researcher to observe participants’ activity through the simulator. Voice feedback was obtained using the Razer Seiren microphone, positioned midway between the participant and monitor for reliable speech detection.

A pre-screen demographic questionnaire was created on Qualtrics. Participants accessed the questionnaire through an anonymous distribution link. The survey directed participants to record prior driving experience, alongside exploring personal disposition to proactive behaviour to assign a fixed LoA in the second condition. Aligning with \citet{li2025framework}’s work on modelling driving-related proactive interactions, dispositional traits similarly outlined in Bateman's Proactive Personality Scale~\cite{bateman1993proactive}, such as propensity to trust and locus of control, are demonstrated to predict engagement with proactive systems when matched with a driver's disposition \cite{li2025framework}. Therefore, proactive disposition for each participant was assessed using Bateman's Proactive Personality Scale~\cite{bateman1993proactive}. 

Their disposition score was normalised within the range [0-1], multiplied by four and rounded to the nearest integer to obtain their LoA. For fixed LoA, 5 participants were assigned LoA 2 (System performs task with user approval), and 14 participants were assigned LoA 3 (System performs task with pending user veto), based on their normalised scores.

\subsubsection{Procedure}


The procedure was divided into three stages: briefing, main experiment, and follow-up questioning. After signing up, participants were sent a confirmation message outlining the study time, location, and researcher contact details. Upon arrival, participants were guided to the study setup. Participants were given the opportunity to adjust their seat, move furniture around the room or change the air-conditioning temperature to ensure comfort. Participants were presented with an online copy of the study information sheet and consent form. Participants were required to complete the demographic questionnaire before engaging with the driving simulator, to ensure key eligibility criteria were met. 

After completing the pre-screen questionnaire, participants were given a verbal walk-through of the experiment. This started with a five-minute explanation of their task to design a proactive IVCA and report their subjective feedback upon finishing each iteration. They were given an introduction to the navigation re-routing scenario. Participants were offered a test run to familiarise themselves with the study apparatus, explore the route, and experience an example intervention from the IVCA.

Participants completed two experiment block sessions, each consisting of a sampling and an optimization phase. Participants completed nine sampling iterations in the trained condition, and seven in the fixed condition, followed by six optimization iterations in both. We did not inform participants that they were undergoing an optimization process or detail how MOBO applied their feedback, following \citet{jansen2025opticarvis}'s argument of not needing in-depth knowledge about an in-vehicle system's operation to evaluate their user experience with it. Participants proceeded through the route, with each iteration ending once they located the designated parking destination and responded to the built-in questionnaire. The researcher was on hand to observe and assist with any technical concerns. Participants were offered a short scheduled break between each block. Participant safety and comfort were monitored throughout with direct observation, alongside preliminary activities including test runs for gradual acclimatisation and reminding participants to inform the on-hand researcher upon immediate discomfort or fatigue. The entire study took approximately 60-75 minutes to complete.

The study finished with a ten to fifteen-minute semi-interview and final debriefing. Participants were given the subject payment form and received 15£ compensation for their time.

The iteration budget for both conditions (nine sampling and six optimization iterations for Trained LoA; seven and six for Fixed LoA) follows prior HITL MOBO research in automotive interface design, where convergence toward Pareto-relevant regions typically occurred within 12–20 iterations~\cite{jansen2025opticarvis}. These limits also reflect practical constraints of VR-based driving studies: longer sequences increase fatigue and degrade the reliability of subjective ratings, whereas shorter sequences risk insufficient model learning. The chosen iteration structure, therefore, balances sample efficiency, participant burden, and expected convergence behavior for subjective measures and multi-objective optimization tasks.

\subsubsection{Ethical Considerations}
 Participants were informed of the potential risks and benefits of the study and encouraged to alert the researcher of any questions. Participation was completely voluntary, and informed consent was provided by each participant before data collection. Anonymity and confidentiality were ensured throughout the study, with data stored on a password-protected, encrypted device viewable by the researchers. All personally identifiable information was omitted during data analysis. Participants were reminded of their freedom to disengage with any portion of the study, take as many breaks as needed, or to opt out of any activities they felt uncomfortable with.

 This study received institutional ethics approval. 

\section{Results}

\subsection{Data Collection}

Participants were gathered using an opportunity sampling approach, allowing easy access to a sample pool that best fit the eligibility criteria.
All personal data, including driving experience and proactive disposition, was solely recorded through the Qualtrics pre-screening questionnaire. Quantitative data, including built-in questionnaire feedback and design parameter values, were obtained using the Unity Bayesian Optimization plugin by Jansen, Colley et al.~\cite{githubGitHubPascalJansenBayesianOptimizationforUnity}. Follow-up interviews were recorded and transcribed using Microsoft Teams. All transcripts underwent a cleaning process prior to analysis. Original recordings were deleted after completing transcription and anonymisation, following institutional guidelines.


\subsection{Data Analysis}

Data analysis consists of two parts: quantitative analysis based on parameter values and questionnaire responses across MOBO iterations, and qualitative analysis to measure participants’ in-depth thoughts. RStudio v2025.05.1-513, R v4.5.1, and RTools v4.5 were used for statistical analysis, alongside a set of enhanced R functions for generating statistical visualizations by \citet{colley2024rcode}.

Semi-structured interviews were conducted to probe into participants’ in-the-moment feelings after experiencing the proactive IVCA. This provided an opportunity to follow up on noteworthy behaviour observed throughout the experiment. \autoref{fig:followup} outlines the set of follow-up questions, to guide the interview process. A reflexive, a-posteriori thematic analysis was used to analyse the transcripts using NVivo 14. As indicated by \citet{braun2019reflecting}, an inductive approach encourages themes to evolve organically to produce a “compelling interpretation”, without pre-defined researcher assumptions of the dataset. 

\subsection{Value Progression Ratings}

\begin{figure*}[ht!]
    \centering
    \begin{subfigure}{0.45\linewidth}
        \includegraphics[width=\linewidth]{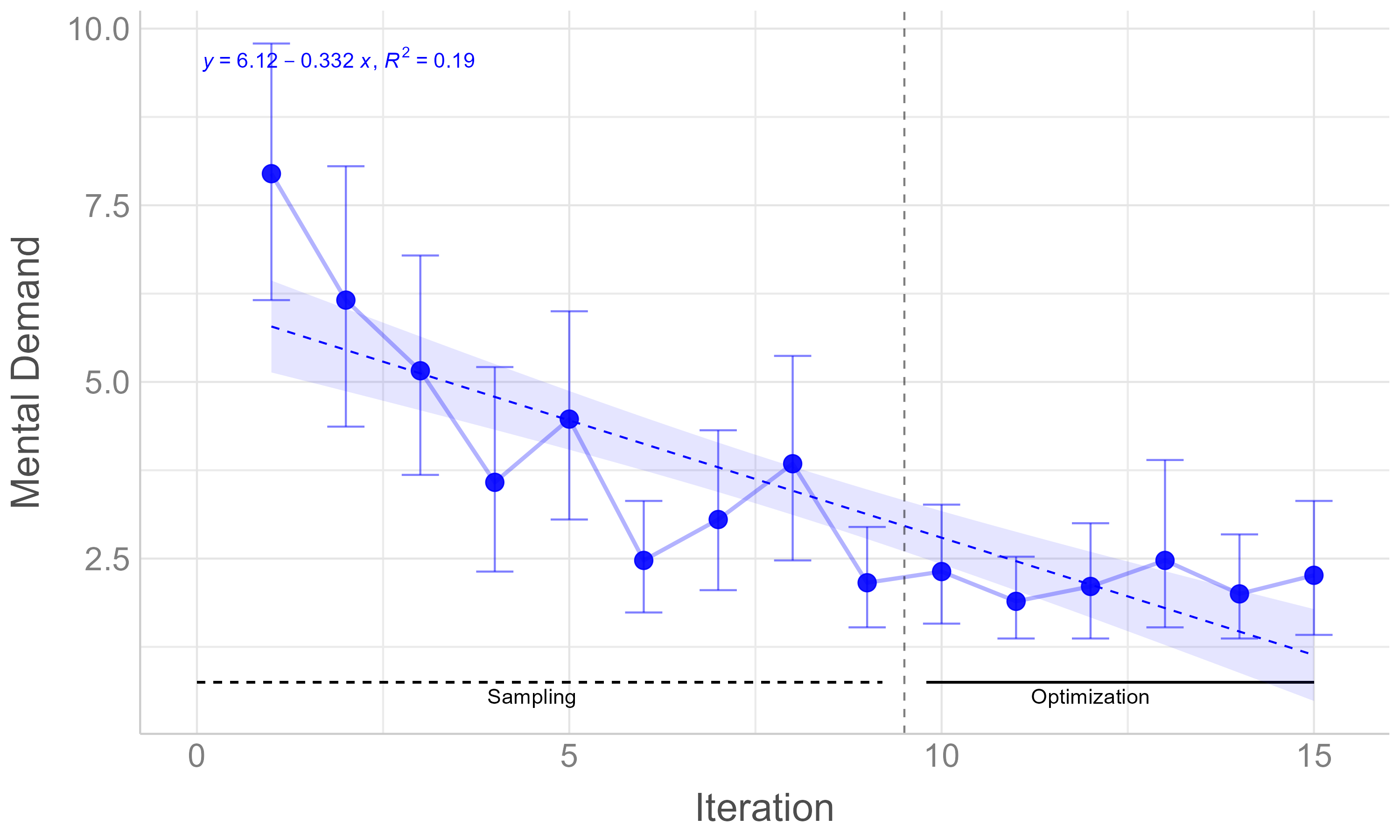}
        \caption{Mental Demand (Condition 1: Trained LoA)}
        \label{fig:c1mental}
    \end{subfigure}
    \hfill
    \begin{subfigure}{0.45\linewidth}
        \includegraphics[width=\linewidth]{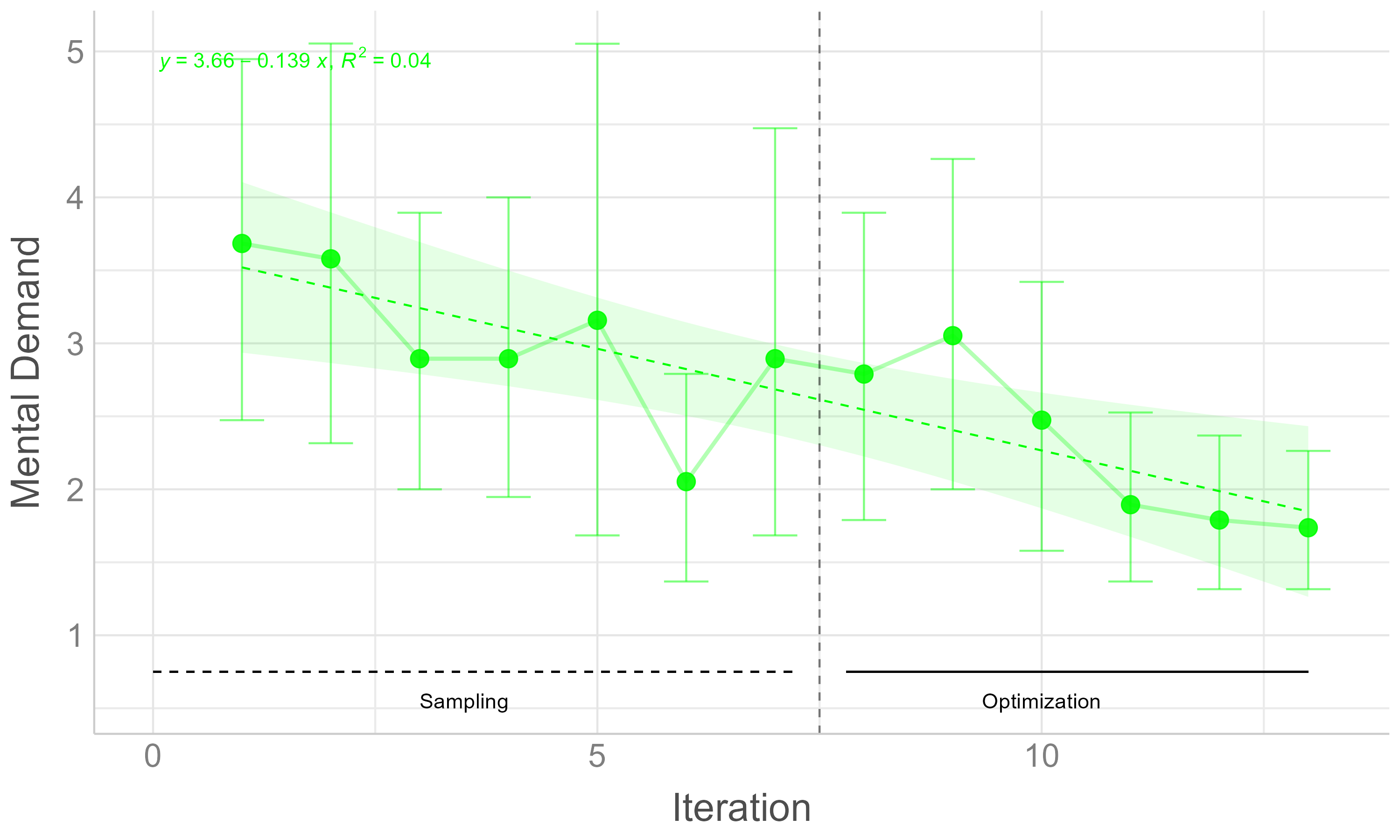}
        \caption{Mental Demand (Condition 2: Fixed LoA)}
        \label{fig:c2mental}
    \end{subfigure}
    \begin{subfigure}{0.45\linewidth}
        \includegraphics[width=\linewidth]{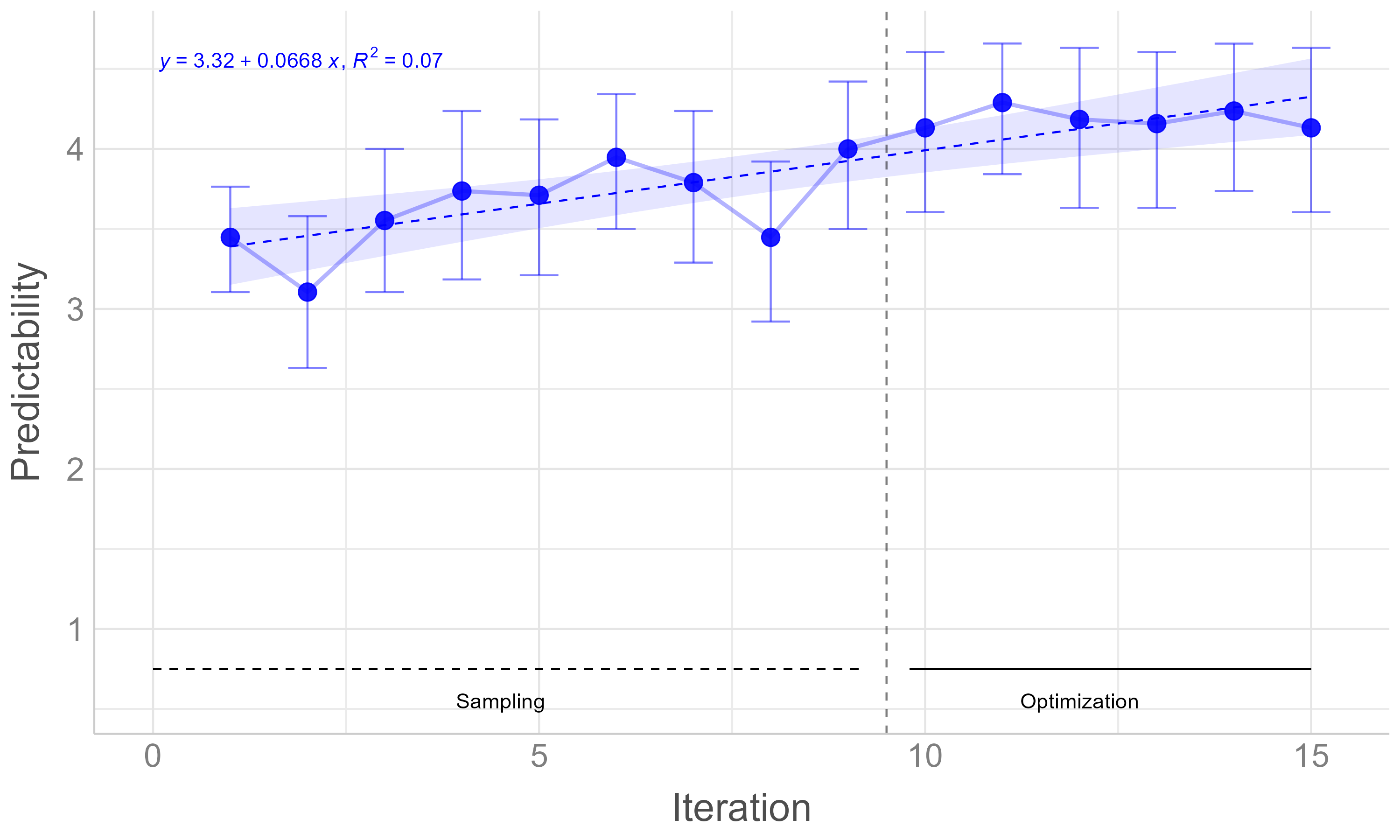}
        \caption{Predictability (Condition 1: Trained LoA)}
        \label{fig:c1pred}
    \end{subfigure}
    \hfill
    \begin{subfigure}{0.45\linewidth}
        \includegraphics[width=\linewidth]{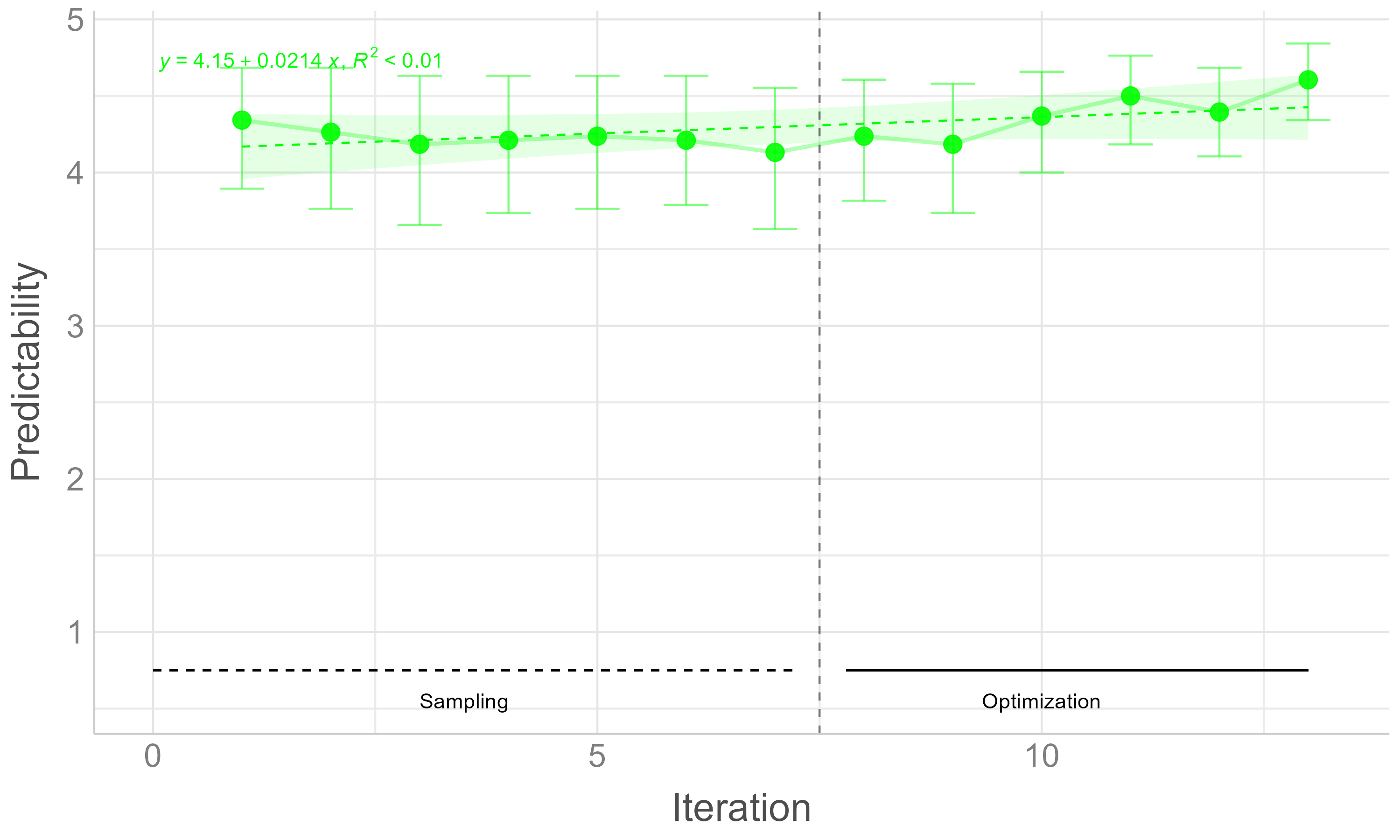}
        \caption{Predictability (Condition 2: Fixed LoA)}
        \label{fig:c2pred}
    \end{subfigure}
    \begin{subfigure}{0.45\linewidth}
        \includegraphics[width=\linewidth]{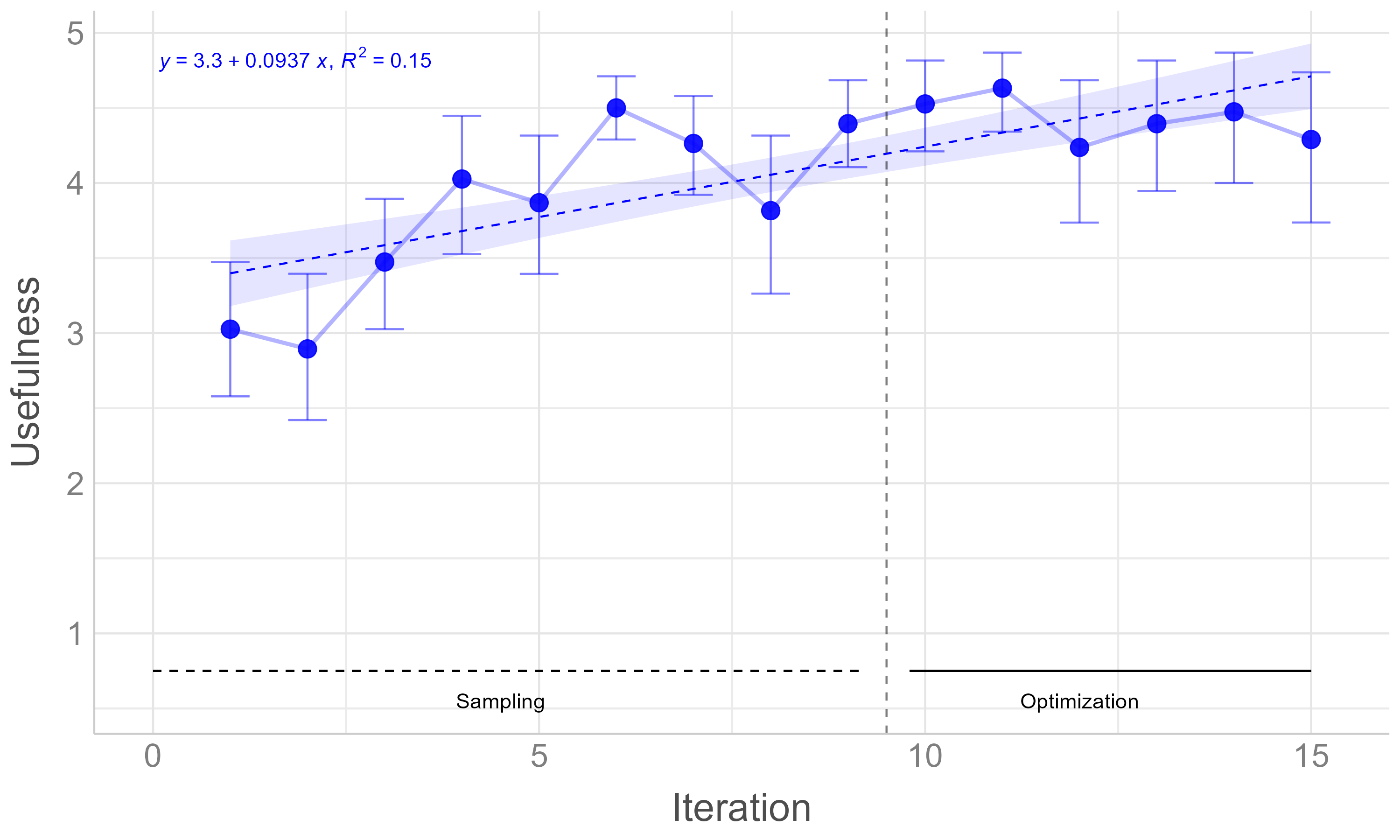}
        \caption{Usefulness (Condition 1: Trained LoA)}
        \label{fig:c1usefulness}
    \end{subfigure}
    \hfill
    \begin{subfigure}{0.45\linewidth}
        \includegraphics[width=\linewidth]{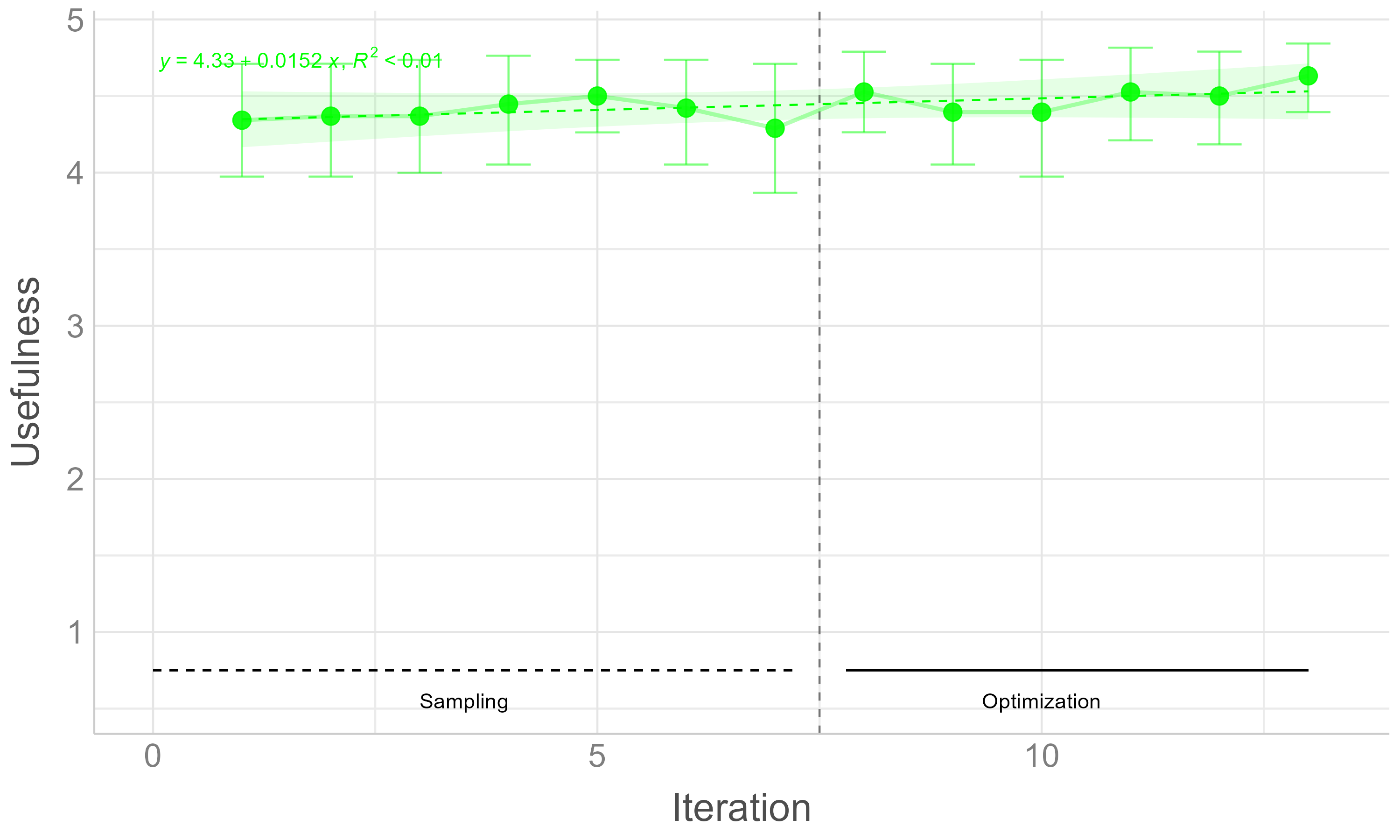}
        \caption{Usefulness (Condition 2: Fixed LoA)}
        \label{fig:c2usefulness}
    \end{subfigure}
    \caption{Progression of mental demand, predictability, and usefulness over MOBO iterations for Trained LoA and Fixed LoA.}
    \Description{Progression of mental demand, predictability, and usefulness over MOBO iterations for Trained LoA (Condition 1) and Fixed LoA (Condition 2).}
    \label{fig:allmeaures}
\end{figure*}

\autoref{fig:allmeaures} outlines the mean value progression of mental demand, predictability, and usefulness in both conditions for all participants who produced at least one design on the Pareto Front, demonstrating that HITL MOBO applied to proactive IVCA personalisation could optimize for design objective values over iterations. Across both conditions, mental demand decreases. The increase in predictability seems to be weaker in Fixed LoA compared to Trained LoA. The increase in usefulness seems to be weaker in Fixed LoA compared to Trained LoA. Interestingly, scores for predictability and usefulness start and remain consistently high when LoA is fixed, compared to trained.

\subsection{Correlation between Design Objectives}

\begin{figure*}[ht!]
    \centering
    \begin{subfigure}{0.45\linewidth}
        \includegraphics[width=\linewidth]{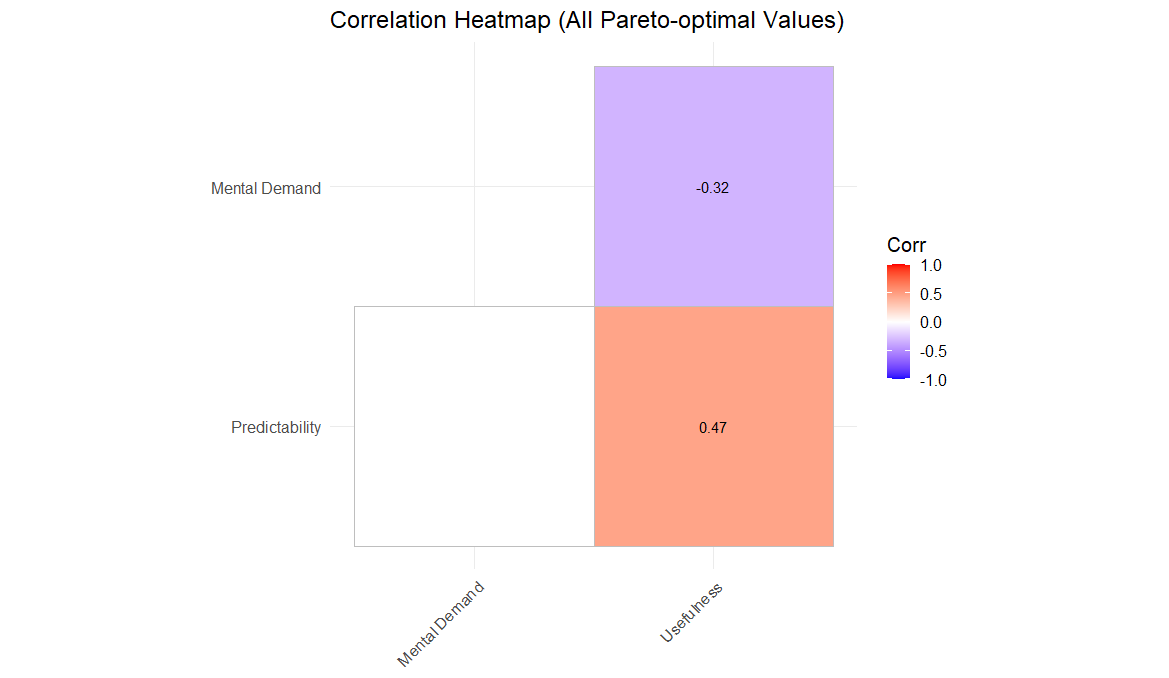}
        \caption{Correlation matrix between mental demand, predictability, and usefulness (All Pareto-optimal solutions)}
        \label{fig:corrpareto}
    \end{subfigure}
    \hfill
    \begin{subfigure}{0.45\linewidth}
        \includegraphics[width=\linewidth]{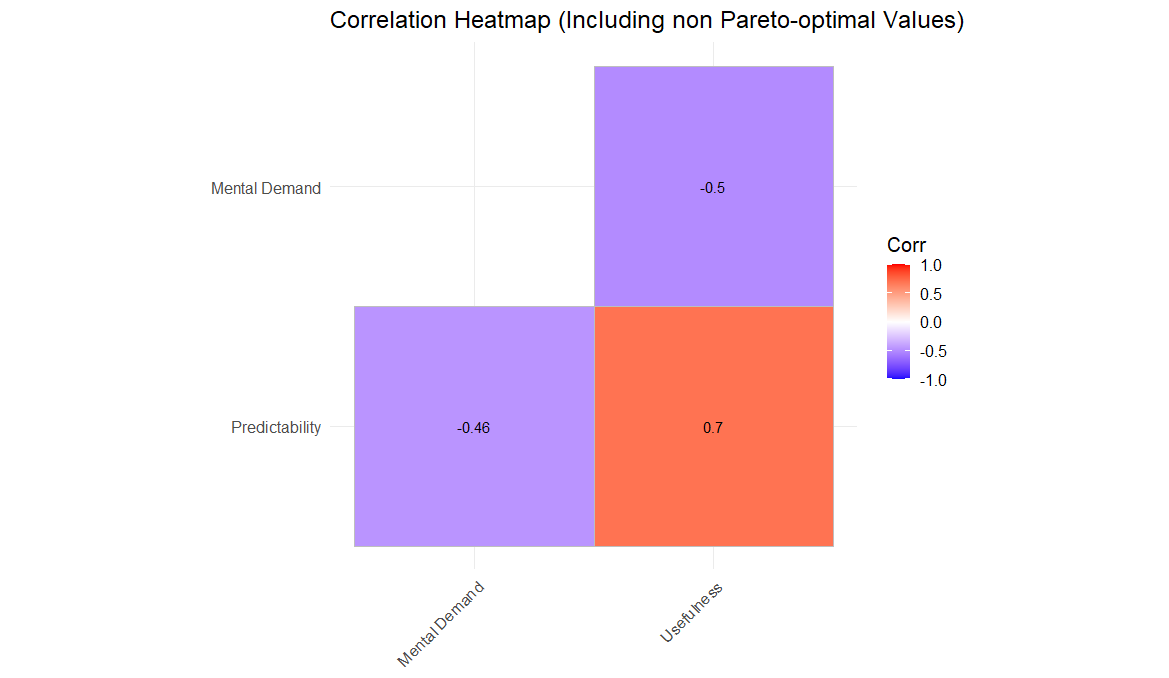}
        \caption{Correlation matrix between mental demand, predictability and usefulness (Including non Pareto-optimal solutions)}
        \label{fig:corrnonpareto}
    \end{subfigure}

        \begin{subfigure}{0.45\linewidth}
        \includegraphics[width=\linewidth]{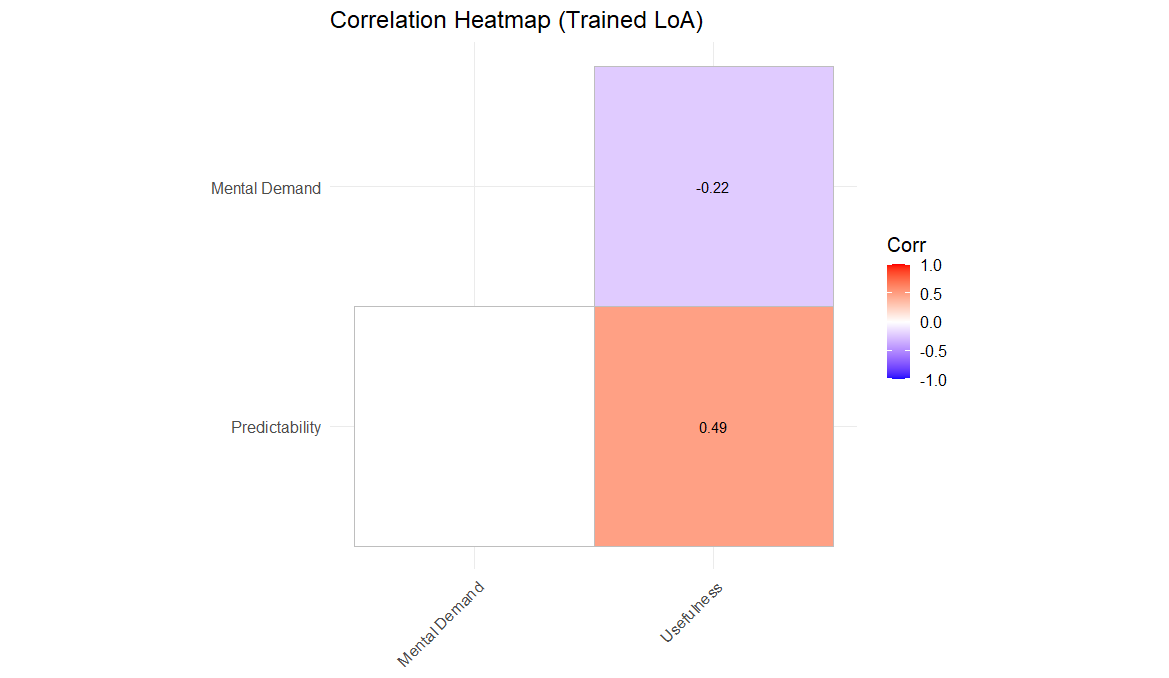}
        \caption{Correlation matrix between mental demand, predictability, and usefulness (Condition 1: Trained LoA)}
        \label{fig:corrcond1}
    \end{subfigure}
    \hfill
    \begin{subfigure}{0.45\linewidth}
        \includegraphics[width=\linewidth]{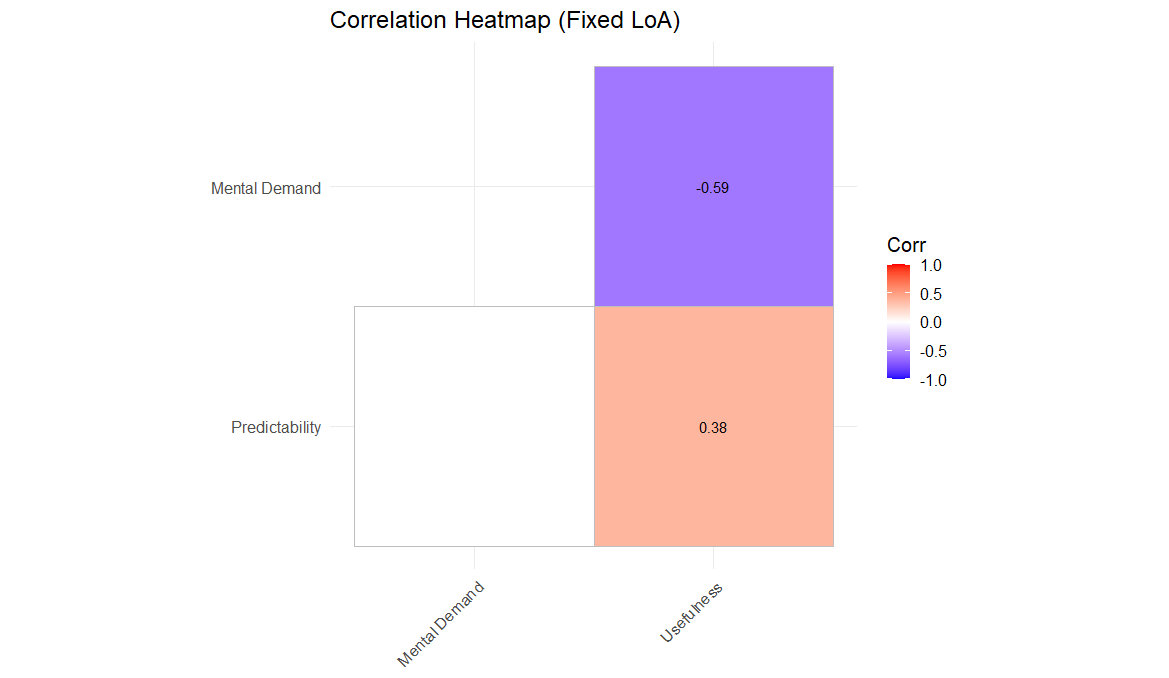}
        \caption{Correlation matrix between mental demand, predictability and usefulness (Condition 2: Fixed LoA)}
        \label{fig:corrcond2}
    \end{subfigure}

    \caption{Correlation between IVCA mental demand, predictability, and usefulness across MOBO iterations.}
      \Description{Correlation between IVCA mental demand, predictabilit,y and usefulness across MOBO iterations, all Pareto-optimal, non-Pareto, condition 1, condition 2.}
    \label{fig:allcorrelation}
\end{figure*}

A pair-wise correlation matrix is calculated using code from Jansen, Colley et al.~\cite{jansen2025opticarvis} to identify the trade-offs between the design objectives and determine how one objective influences another. \autoref{fig:allcorrelation} provides a visual representation using the Pearson Correlation Coefficient~\cite{sedgwick2012pearson}. Across all 214 Pareto-optimal values in both conditions, predictability and usefulness had a strong positive correlation (r = 0.47), meaning that improvements in one led to an increase in the other. Mental demand and usefulness had a strong negative correlation (r = -0.32), suggesting that lower mental demand is associated with increased usefulness. There was no significant correlation between mental demand and predictability.

Predictability and usefulness had a strong positive correlation (r = 0.49) when LoA was trained. There was no significant correlation between mental demand and predictability. When LoA was fixed, there was a strong positive correlation between predictability and usefulness (r = 0.38), and a strong negative correlation between mental demand and usefulness (r = -0.59).

Much greater significant correlations exist when considering all parameter values (19 * 28 iterations = 532 values). Particularly, there remains a strong positive correlation between predictability and usefulness (r = 0.7). Mental demand and predictability had a strong negative correlation (r = -0.46).

\subsection{Pareto Front Design Parameter Set}

\begin{table*}[ht!]
\centering

\caption{Mean, median, standard deviation (SD), interquartile range (IQR), minimum, and maximum for all Pareto-optimal design parameter values, split into both conditions.}
\resizebox{\textwidth}{!}{%
\begin{tabular}{lcccccc|cccccc}
\toprule
& \multicolumn{6}{c}{\textbf{Trained LoA}} & \multicolumn{6}{c}{\textbf{Fixed LoA}} \\
\cmidrule(lr){2-7} \cmidrule(lr){8-13}
\textbf{Parameter} & \textbf{Mean} & \textbf{Median} & \textbf{SD} & \textbf{IQR} & \textbf{Min} & \textbf{Max} & \textbf{Mean} & \textbf{Median} & \textbf{SD} & \textbf{IQR} & \textbf{Min} & \textbf{Max} \\
\midrule
$p_1$: Interior Lighting Glow & 0.371 & 0.330 & 0.267 & 0.315 & 0 & 1 & 0.411 & 0.390 & 0.336 & 0.493 & 0 & 1 \\
$p_2$: Auditory Alert Volume & 0.569 & 0.600 & 0.237 & 0.390 & 0.1 & 1 & 0.570 & 0.480 & 0.306 & 0.590 & 0.1 & 1 \\
$p_3$: Symbol Transparency & 0.530 & 0.565 & 0.231 & 0.410 & 0.1 & 1 & 0.613 & 0.725 & 0.310 & 0.52 & 0.1 & 1 \\
$p_4$: Level of Autonomy (LoA) & 0.634 & 0.690 & 0.267 & 0.476 & 0 & 1 & - & - & - & - & - & - \\
\bottomrule
\end{tabular}
}

\label{table:paretofrontvalues}
\end{table*}

\begin{figure*}[ht!]
  \centering
  \includegraphics[width=\linewidth]{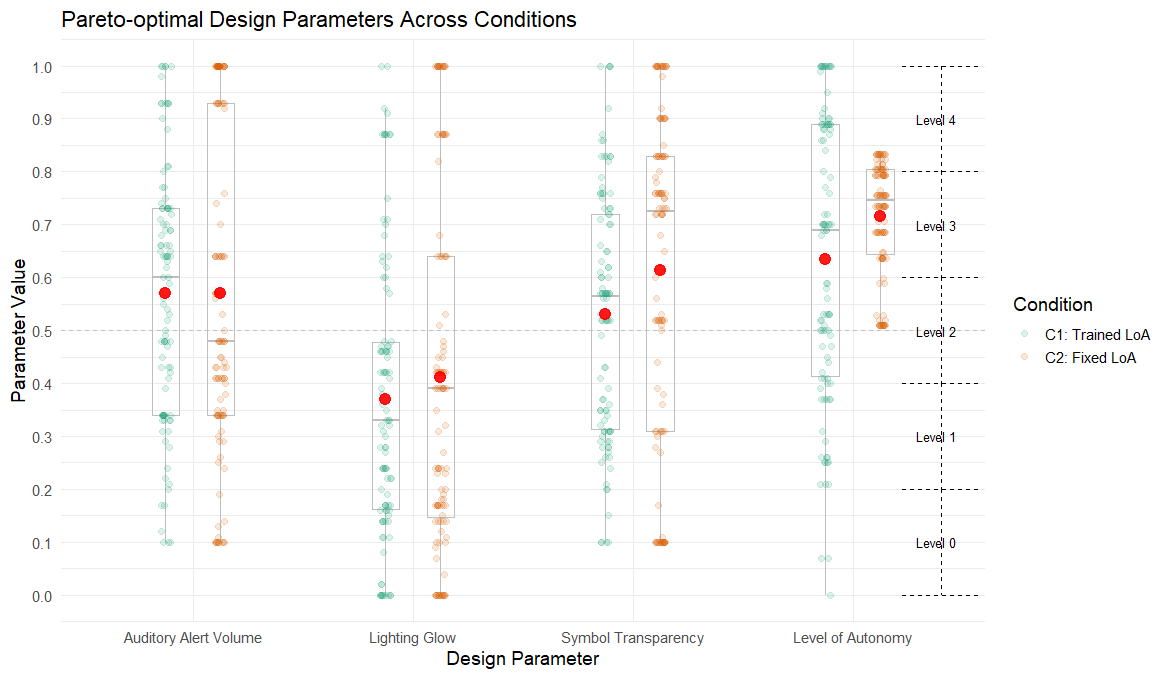}
  \caption{Scatter distribution and box plots of all Pareto-optimal design parameter values, split into both conditions.}
   \Description{Scatter distribution and box plots of all Pareto-optimal design parameter values, split into both conditions.}
  \label{fig:paretoScatterPlot}
\end{figure*}

\begin{figure*}[ht!]
    \centering

      \begin{subfigure}{0.49\linewidth}
    \centering
    \includegraphics[width=\linewidth]{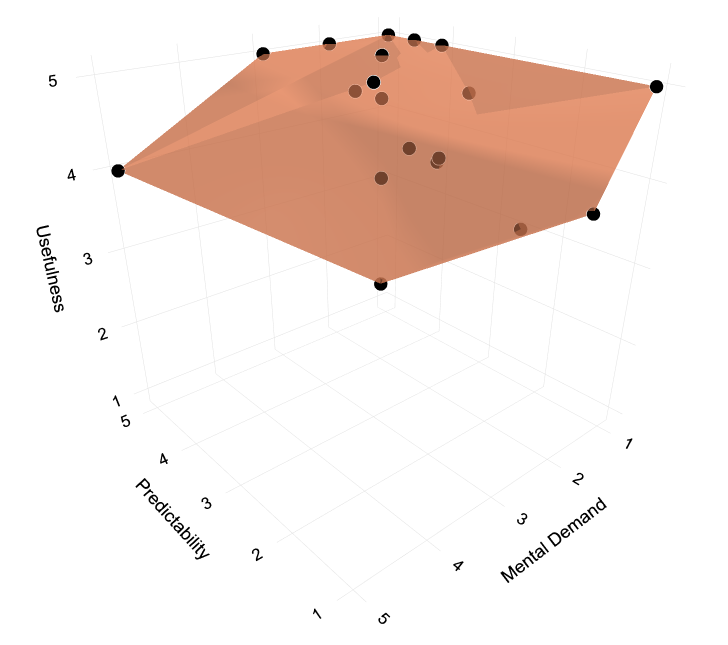}
    \caption{Pareto Front Distribution (Trained Condition)}
    \label{fig:pareto3dTrained}
\end{subfigure}
     \hfill
\begin{subfigure}{0.49\linewidth}
    \centering
    \includegraphics[width=\linewidth]{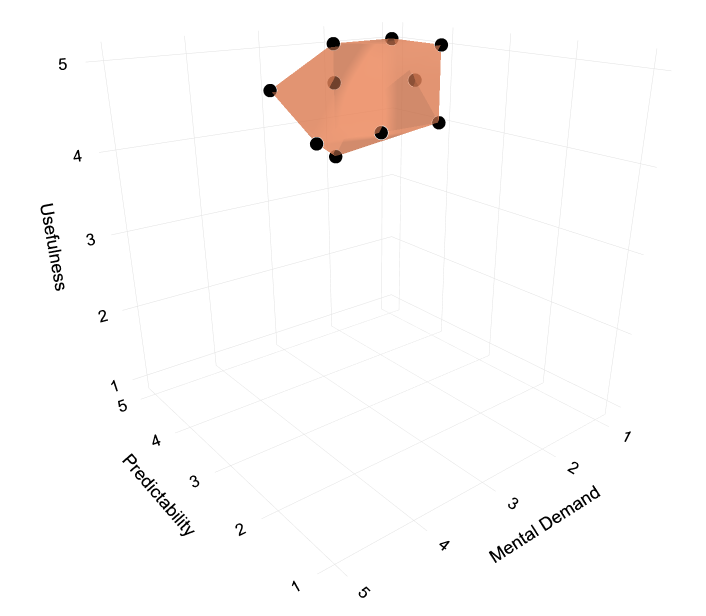}
    \caption{Pareto Front Distribution (Fixed Condition)}
    \label{fig:pareto3dFixed}
\end{subfigure}
    \caption{Three-dimensional scatter distribution of Pareto-optimal solutions, for all design objectives.}
    \Description{Three-dimensional scatter distribution of Pareto-optimal solutions, for all design objectives.}
    \label{fig:paretofront}
\end{figure*}

 \autoref{fig:paretofront} visualises the three-dimensional scatter distribution of all Pareto-optimal solutions for all objectives. Combined, these solutions represent the Pareto Front - the most effective trade-offs when balancing the design objectives~\cite{jansen2025opticarvis}. All points in the set are non-dominated. Notably, all Pareto-optimal designs are close to the design space boundaries.

 \autoref{table:paretofrontvalues} and \autoref{fig:paretoScatterPlot} visualise the final parameter sets across both conditions, illustrating the mean, median, standard deviation, IQR, minimum, and maximum parameter values on the Pareto Front. Interestingly, the average trained LoA coincides with the average score for participants' self-reported proactive disposition after normalising.

\subsection{Qualitative Analysis}

After completing both conditions, all participants provided open-ended feedback through a set of semi-structured interview questions. A deductive thematic analysis was used to analyse this feedback. The findings are divided into four overarching themes.

\subsubsection{General Impressions towards Multi-Objective Bayesian Optimization}

Participants were impressed by MOBO's ability to adapt the proactive IVCA's design. Notably, P1 likened the experience to an evolutionary or learning process: \textit{“Like in the first several travels [...], I don't know what he's gonna do [...] but after several trials on learning, it evolves somehow”}. P10, a Human-Computer Interaction design expert, sees value in the scalability of MOBO: \textit{“It’s quite helpful [...] in terms of like you can quickly deploy this to many users and you don’t have to manually adjust it [...]”}.


\subsubsection{Proactive Intervention Preferences}

Participants saw potential in future IVCAs behaving proactively, disclosing on-road scenarios where they could see themselves interacting with it. For instance, P8 delved into challenges with populated destinations, particularly when searching for available parking. However, after engaging with ProVoice, they emphasised the value of the assistant’s instructions for managing potential congestion: \textit{“It wouldn't be a case of you having to constantly roam around. Especially for like big places [...] that always get really populated. It'd probably be way easier compared to massive build-ups”}.

Conversely, participants disclosed scenarios where such proactivity may be undesirable. Two participants share a common desire towards the intervention, aligning with their navigation goals in the moment. Without alternative choices for re-routing, they felt the proactive intervention would constrain their options: \textit{“Say I'm the user and my preference may be that I want it to automatically just take me to naturally the best place it thinks it is. But what if I'm someone who is more cost saving? What if I don't actually want to pay the £2 per hour to park? What if I actually wanted some free street parking somewhere along?”} (P2). Similarly, P12 speculates alternative ideas for re-routing should they reject the initial intervention: \textit{"Would it have come back and ask you why? Or would it give me another alternative? [...] Maybe if you were on a longer stretch of road, it might give you the option - do you want to choose this one that's 50 metres or 200 yards ahead, or do you want to travel one mile ahead?"}.

\subsubsection{Personal Preferences Towards Level of Autonomy}

Participants held varying beliefs on the LoA, which felt suitable to them. P6 disclosed a preference towards lower LoA due to the independent control offered: \textit{“I think it's better when it asks for your response because you might not want to go to that parking spot or you might want to just stop right outside the building”}. Conversely, P2 drew attention to this control as a potential safety risk due to diverting driver focus: \textit{“Having to listen to the proactive assistant and then answer may or may not be distracting for certain people [...] that can actually be, from my perspective, a little dangerous because you are taking attention away from the driver to ask them a question and in which you will then possibly change directions or navigation that they were not expecting.”}. 

\subsubsection{Limitations of Proactive Intervention}

Participants disclosed certain frustrations about interacting with the IVCA, especially considering the way that information was conveyed throughout the experience. Four participants reported fixating on the voice output and auditory alert, to the point of missing the differences in visual UI cues. When probed further, P8 attributed this to focusing on the road ahead: \textit{“I mean mostly I was focusing on the route [...] 'cause the road is more unpredictable and there'll be random lorries that come through, you know?”}. P2 mentioned their worries about the amount of surveillance required for the IVCA to function: \textit{“I think from a skeptic side, how much information is this programme and assistant listening to gather the needs, to actually be a good navigation assistant”}.

\section{Discussion} 
A within-subjects study integrated MOBO to optimize ratings of mental demand, predictability, and usefulness while customising proactive IVCA variants. The following section covers (1) findings on HITL MOBO for proactive IVCA personalisation in relation to prior work, (2) implications for different stakeholders, and (3) wider ethical consequences for computational design techniques.

\subsection{Computational Techniques for Proactive Intervention Design}

Addressing RQ1, proactive intervention produced by HITL MOBO reduced mental demand, while increasing predictability and usefulness, across both conditions (see \autoref{fig:allmeaures}). \autoref{fig:allcorrelation} highlights a strong positive correlation between predictability and usefulness, and a strong negative correlation between mental demand and predictability for all Pareto-optimal values across both experiment conditions. Contrasting with user-led in-vehicle personalisation techniques explored by \citet{normark2015design}, HITL MOBO encourages systematic exploration through the design space to discover optimal IVCA variants. Aligning with prior work into HITL MOBO as a design technique~\cite{jansen2025opticarvis, colley2025improving}, these findings reaffirm the success of optimization-driven approaches for personalising automotive functionality, while embracing differences in driver heterogeneity~\cite{yarlagadda2022heterogeneity}. 

However, the success of HITL MOBO raises concerns with regard to the designer's relevance during personalisation. While responsible for defining the design space and objective functions, HITL MOBO could allow the designer to shift the responsibility of uncovering optimal IVCA variants, due to emphasis on end-user feedback. Despite this, the algorithmic approach is regarded as scalable, especially from the expert designer's viewpoint of identifying variants for each end-user. Furthermore, addressing RQ2, mean driver ratings towards predictability and usefulness of the proactive IVCA remained consistently high, and mental demand low, when LoA was fixed based on Bateman's Proactive Personality Scale~\cite{bateman1993proactive}, compared to training as a design parameter during MOBO (see \autoref{fig:allmeaures}). One potential explanation for these ratings is that fixing LoA from the offset aligned the system with the driver's expectations for proactive behaviour during early iterations. In contrast, training LoA required several iterations to adjust and match expectations. This resulted in a higher baseline perceived predictability and usefulness, coinciding with their dispositional traits, alongside a smaller distribution of Pareto-optimal solutions (see \autoref{fig:paretofront}). These results indicate value in conducting additional user-centred design activities like preliminary questionnaires, to uncover driver needs and preference \textbf{before} running MOBO. This could position the designer as a collaborator throughout optimization, allowing them to target specific regions in the design space and create variants that quickly converge to optimal scores, rather than complete delegation to HITL MOBO. By running MOBO with a relatively small study sample, designers can obtain Pareto-optimal parameter values to derive and test one-size-fits-all intervention variants, for example, on-road scenarios, prior to wider consumer release (see \autoref{fig:paretoScatterPlot}). This can establish a starting point for end-users, from which they provide additional HITL feedback throughout driving journeys to further design their in-vehicle experience. In practice, both designer and end-user should collaborate to achieve in-depth exploration through design spaces, with HITL MOBO as a key tool to drive the search.

The use of MOBO shapes how the proactive IVCA design space is explored. Rather than sampling designs uniformly, the optimizer emphasizes regions where user preferences are uncertain or where meaningful trade-offs between usefulness, predictability, and mental demand may arise. We argue that this guided search is appropriate for subjective, iterative evaluations because the progression of ratings across iterations (see \autoref{fig:allmeaures}) reflects how participants differentiate between successive design variants.
However, repeated questioning on the same route, changing situational cues, and growing familiarity with the driving task can all independently shift ratings, even without optimization. Such effects introduce noise and learning dynamics that limit the degree to which convergence of design-objective values toward their optimum can be expected within a single-session study. We therefore interpret the observed rating trends as approximate convergence: the limited iteration budget (set to avoid VR fatigue) cannot fully counteract subjective variability, but it can still guide the optimization toward regions of the design space where Pareto-optimal designs are plausible. Similar rating convergence has been reported in prior HITL optimization of subjective design objectives~\cite{jansen2025opticarvis, colley2025improving}. Thus, we argue that MOBO achieves its practical goal in this context: supporting efficient, noise-aware adaptation rather than identifying a definitive optimum. Reaching such an optimum may be possible with substantially more HITL iterations; however, doing so would require considerably more user time and effort.

Another concern regarding the HITL optimization approach is that users may assume that designs should improve over time (forming a self-fulfilling prophecy). However, we observed that the iteration-to-iteration rating changes were not monotonic, and participants provided divergent responses to different design variants, which may suggest that evaluations were driven by the presented IVCA behaviour rather than by a general belief that later iterations must be better. Moreover, the strongest shifts in usefulness, predictability, and mental demand occurred in iterations where the optimizer selected substantially different design variants, indicating sensitivity to changes in the underlying design parameters. Still, the way participants are primed about the optimization process itself, such as explicitly stating that the system is optimizing a design versus withholding this information, may substantially shape their expectations. Future work should explore these priming effects when employing computational techniques for design.

\subsection{Safety Considerations for Personalised Proactive Intervention}

Participants identified future on-road scenarios where such proactive functionality would be valuable, such as navigating congested areas and conversing with the IVCA to discover alternative re-routing options. However, an important gap emerged regarding potential threats to driver safety while engaging with proactive functionality. Scepticism was held towards the data that a real IVCA might need to collect, reflecting wider worries of data privacy and misuse previously identified by \citet{reicherts2021may}. This further calls for the assistant to be transparent and explainable in its data use. Such approaches to uphold data transparency might include requesting prior consent before starting a journey, or providing methods to opt out of transmitting end-user feedback to OEMs. Safeguards for protecting data privacy will be crucial towards strengthening future engagement with the IVCA, given that frequent surveillance of the vehicle may be necessary to learn opportune intervention points. Speculating into potential surveillance and feedback methods, recent initiatives by Garmin have supported wearable technology to collect implicit telemetric data, including driver pulse and heart rate, for adapting the vehicle's infotainment system to enhance well-being (see Mercedes-Benz Energising~\cite{energising}). Such metrics could be beneficial to determine suitable timing, particularly when deciding appropriate LoA according to driver focus and state. Built-in sensors, such as pressure points around the steering wheel, could track driver focus in real-time based on grip force and hand placement, without requiring external tracking tools. However, trade-off between privacy and functionality could arise if the driver restricts access to their feedback, limiting the IVCA's potential to align with end-user traits.

Despite lowered mental demand across both experiment conditions in \autoref{fig:allmeaures}, participants referred to the IVCA's utterances as diverting attention from the road, potentially overwhelming or distracting from the primary driving task. In particular, P2 held concerns towards Level 2 (System performs task with user approval) as dangerous due to prompting a potential unexpected response. 
In addition, when looking at all Pareto-optimal solutions when LoA is trained, average LoA converged to level 3 (System performs task with pending user veto), matching 14 participants' self-reported proactive disposition for Fixed LoA. Combined with prior work on intervention timing~\cite{semmens2019now}, such feedback underscores the importance of modelling intervention design that accounts for driver availability to communicate, to mitigate any impacts to stress or workload. The driving environment is dynamic in that drivers must perform instantaneous, non-routine actions such as rapid adjustments to speed or acceleration~\cite{wang2015level}. Thus, a driver's capacity to interact with proactive intervention may depend on their current on-road situation. Alongside identifying optimal points of intervention~\cite{semmens2019now}, a proactive IVCA could benefit from adjusting LoA based on the situation. One method to achieve this, similar to \citet{semmens2019now} and explored during the parking scenario, might be through gathering driver feedback at a safe point. Through this, the IVCA can learn optimal interventions for each on-road scenario over time. For instance, a driver may opt for proactive intervention during calm moments, such as while stationary at a traffic light, to have a different level of user agency in comparison to volatile situations like high-speed driving.


\subsection{Design Implications}
\subsubsection{Prioritising Auditory-first Driving Interventions}
Four participants disregarded visual cues conveyed by the IVCA, including the dashboard glow and infotainment symbol. Building on \citet{sodnik2008user}, a competing attention between the demands of the road and visual in-vehicle interfaces leads to reduced willingness to engage with the visual cues while focused on the driving task. Voice-based modalities, including voice output and auditory alerts, could be appropriate for supporting access to in-vehicle services amidst the visual demands of the road.

\subsubsection{Aligning Autonomy with Granular Driver Preferences/On-road Situations}
Beyond design modalities, the LoA must adapt to the driver's specific intent for the current journey. Qualitative feedback revealed that rigid autonomy can conflict with user goals; for instance, P2 noted that automated re-routing (Level 4) might bypass their personal preference for "cost saving" or "street parking". Similarly, P6 expressed a preference for lower autonomy (Level 1 or 2) to maintain the freedom to "stop right outside the building" rather than being forced into a parking destination. These findings indicate that autonomy should be a situational negotiation based on the driving context. In volatile scenarios like high-speed driving, the system might shift towards higher autonomy (Level 3 or 4) to minimise conversational distraction, while reserving lower autonomy (Level 1 or 2) for situations where the driver has the capacity to negotiate suggestions such as when stationary in traffic.

\subsection{Limitations and Future Work}

The present study encountered several limitations to consider. First, ProVoice focused on implementing MOBO using a ‘cold-start’ approach. This relies on initialising parameter values at random~\cite{jansen2025opticarvis}. While cold-start MOBO was successful in optimizing for the design objectives across iterations, future research may consider evaluating alternative approaches to MOBO to promote in-depth exploration through the design space. This might include user or expert-informed ‘warm-start’ approaches to adjust initial values according to pre-created IVCA variants~\cite{jansen2025opticarvis}.

While not directly informed that they were going through an optimization process, participants’ own awareness of going through a HITL MOBO process and providing iterative feedback means that they might expect to report increasing subjective scores across the experiment, independent of actual design parameter changes. This could result in a self-fulfilling prophecy feedback loop. While not employed during the study, future work could benefit from implementing control trials, where the system produces random IVCA variants during certain iterations, to see if subjective values continue to improve. Repeated experiments with the same participants across different optimization processes could determine whether MOBO converges on the same result.

While focusing on a smaller sample size allowed for detailed observations and questioning, any extreme responses, including outliers or disengaged participants, may distort the quantitative findings. Future work might consider gathering a wider range of participants from different recruitment channels while implementing alternative approaches beyond opportunity sampling to eliminate selection bias based on participant convenience. There is also potential to group participants with varying ranges of driving experience, age, and technological expertise, for further comparison between feedback on IVCA design preferences. Furthermore, participants were distributed between LoA 2 and 3 in the fixed experiment condition, with no participants reporting a proactive disposition to assign them into level 0, 1, or 4 after rounding. Outreaching to participants with different reported proactive dispositions may allow for nuanced comparisons between each LoA.

The study defines proactivity using LoA as the main design parameter, focusing on one scenario with specific timing and relevance. This opens up the potential to apply optimization techniques in future studies to determine opportune points of interaction across journeys, beyond the design of the intervention itself.

Despite measures to prevent simulator sickness, including realistic driving controls, direct monitoring, and running the study in a well-lit, well-ventilated room, participants would sometimes alert the researcher to remove the VR headset for a short break. However, this seemed to vary, with many participants completing all iterations without removing the headset. Every participant completed the experiment. Future studies should accommodate any individual sensitivities that participants might have when faced with sensory conflict and employ techniques to control and limit potential nausea. This could be through incorporating frequent scheduled study breaks or including prior VR experience as a key eligibility criterion, to ensure a baseline level of prior exposure.

Future work should also prioritize conducting evaluations in real vehicles under dynamic driving conditions, for instance using XR-OOM~\cite{xroom}, Portobello~\cite{portobello}, or PassengXR~\cite{PassengXR}. Where this is not technically feasible, a vehicle motion simulator can serve as an alternative (e.g., \cite{colley2021swivr, 10.1145/3543174.3545252}).

Repeating a consecutive driving journey in short succession can introduce learning effects as a confounding variable, due to increased experience with the simulator and IVCA. Additional research might consider slight variations in the driving scenario during iterations. This might be through adjusting the final destination, or a deviation in the route to introduce new manoeuvres, such as a new turn, roundabout exit, or mirroring the entire journey.

\section{Conclusion}
In conclusion, this project introduces ProVoice, a novel driving simulator prototype that evaluates HITL MOBO as a computational design technique for proactive IVCA personalisation. A VR driving simulator study with 19 participants validates the approach, discovering reports of lowered mental demand, alongside increased predictability and usefulness while interacting with the proactive IVCA across both experiment conditions. While the prototype considers four design parameters (LoA, symbol transparency, alert volume, and interior glow) and three objectives (mental demand, predictability, and usefulness), ProVoice can be extended to include additional design elements, facilitating broad exploration of personalised intervention design. The source code for proactive intervention logic and MOBO implementation, as seen in ProVoice, is publicly available. This includes a README for setting up the asset in other projects, alongside a prototype demonstration and instructions on how to extend in future research.

\section*{Open Science}
The open-source code for implementing proactive intervention, as seen in ProVoice, is released online (see \href{https://github.com/JSusak/ProVoiceProactiveIntervention/}{ProVoiceProactiveIntervention Repository}). This repository includes a README for setting up the asset in other Unity projects, alongside an installation tutorial. Quantitative participant data, including questionnaire feedback and parameter values obtained during HITL MOBO, and analysis code are released online (see \href{https://github.com/JSusak/ProVoiceData}{ProVoiceProactiveIntervention Data}).

\begin{acks}
We thank all study participants.
\end{acks}

\bibliographystyle{ACM-Reference-Format}
\bibliography{main}

\appendix

\section{Materials}

\subsection{Proactive Driving Scenario}
\label{appendix:intervention}

\textit{Imagine that you have already been driving for one hour now on your way to a casual business meeting with your colleagues at a coffee shop. You will arrive in a few minutes but the shop has no available on-street parking. Your goal is to locate and reach a nearby parking lot (marked in yellow) to attend the meeting in good time.}

\subsection{Assistant Responses per LoA}
\label{appendix:responses}

\begin{itemize}
    \item \textbf{Level 0 [User performs task without system support]:} \textit{"You are about to arrive at your destination. Your destination will be on the left."}
    \item \textbf{Level 1 [User performs task with system support]:} \textit{"You are about to arrive at your destination, but there doesn't seem to be any parking. You may need to search for a nearby parking lot."} -> "Search/find" -> \textit{"Okay, I've found a nearby parking lot 0.2 miles from the destination. I'm redirecting you there now."}
    \item \textbf{Level 2 [System performs task with user approval]:} \textit{"You are about to arrive at your destination. There seems to be no on-street parking available. Should I already look for a parking spot nearby?"} -> "Yes" -> \textit{"Okay, I've redirected your navigation to the closest parking lot, 0.2 miles away with a price of £2 an hour."}
    \item \textbf{Level 3 [System performs task with pending user veto]:} \textit{“You are about to arrive at your destination. There seems to be no on-street parking available. I will redirect your navigation to a parking lot 0.2 miles from the destination, with a price of £2 an hour. Please interrupt if you would like to change or cancel.”}

    \item \textbf{Level 4 [System performs tasks automatically/no user intervention]:} \textit{"Your destination has limited parking. I have already redirected your navigation to the closest parking lot 0.2 miles from the destination, with a price of £2 an hour."}
\end{itemize}



\begin{figure}[ht!]
    \centering
     \includegraphics[width=0.5\textwidth]{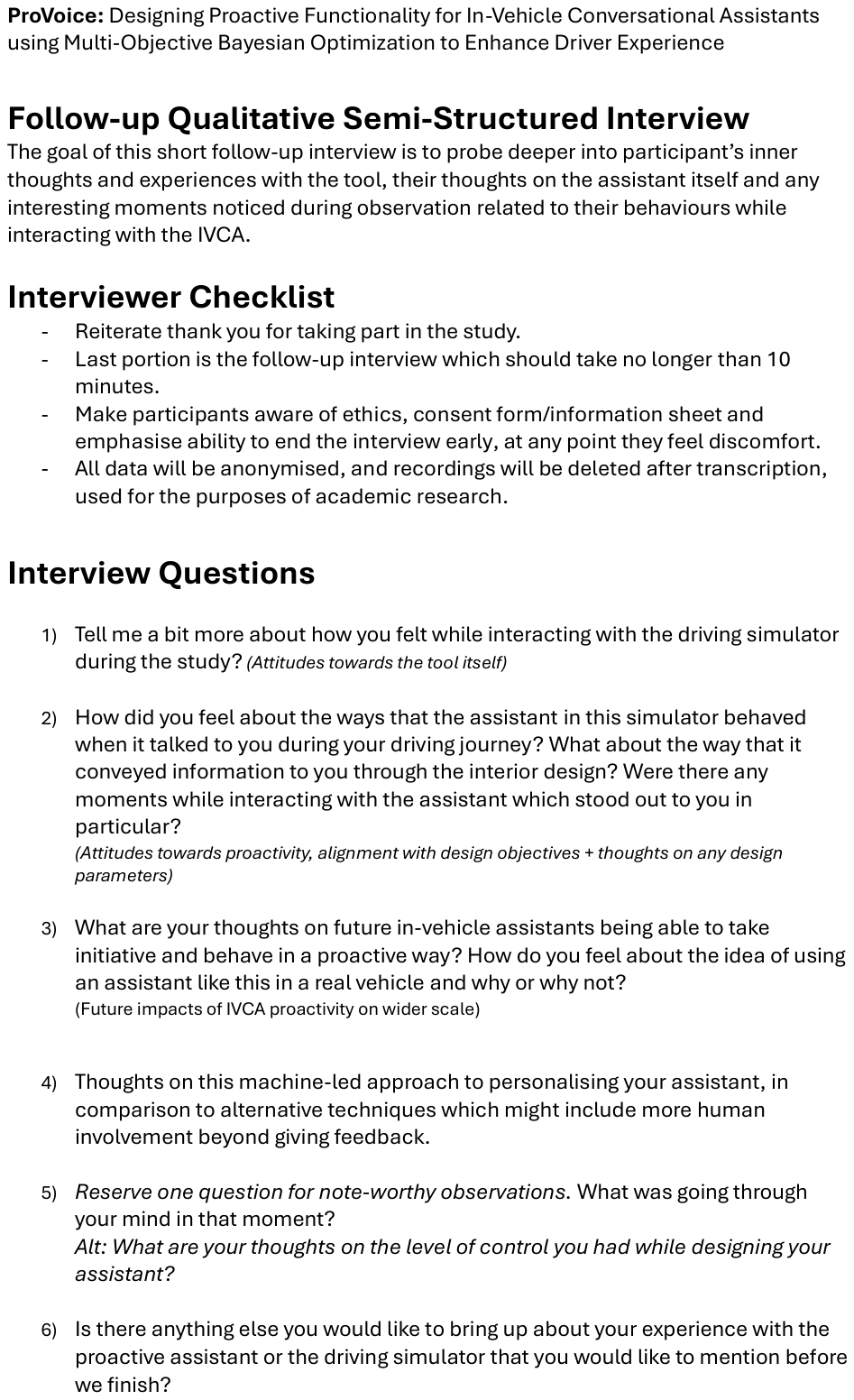}
     \caption{Follow-up Interview Questions.}
     \Description{Follow-up Interview Questions.}
     \label{fig:followup}
 \end{figure}

\begin{figure}
    \centering
     \includegraphics[width=0.5\textwidth]{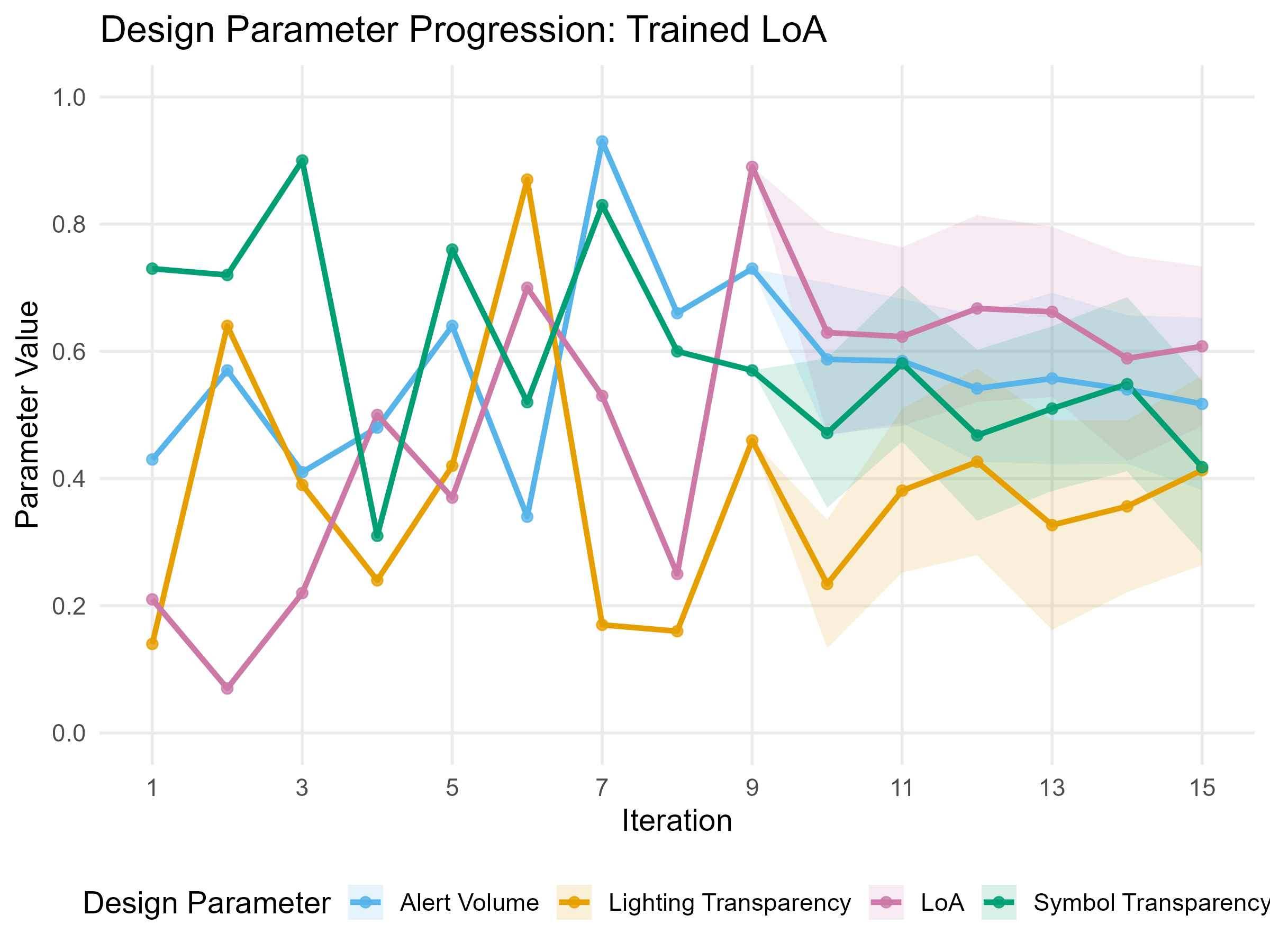}
     \caption{Mean design parameter progression (Trained Condition)}
     \Description{Mean design parameter progression (Trained Condition).}
     \label{fig:parameterProgressionTrained}
 \end{figure}

 \begin{figure}
    \centering
     \includegraphics[width=0.5\textwidth]{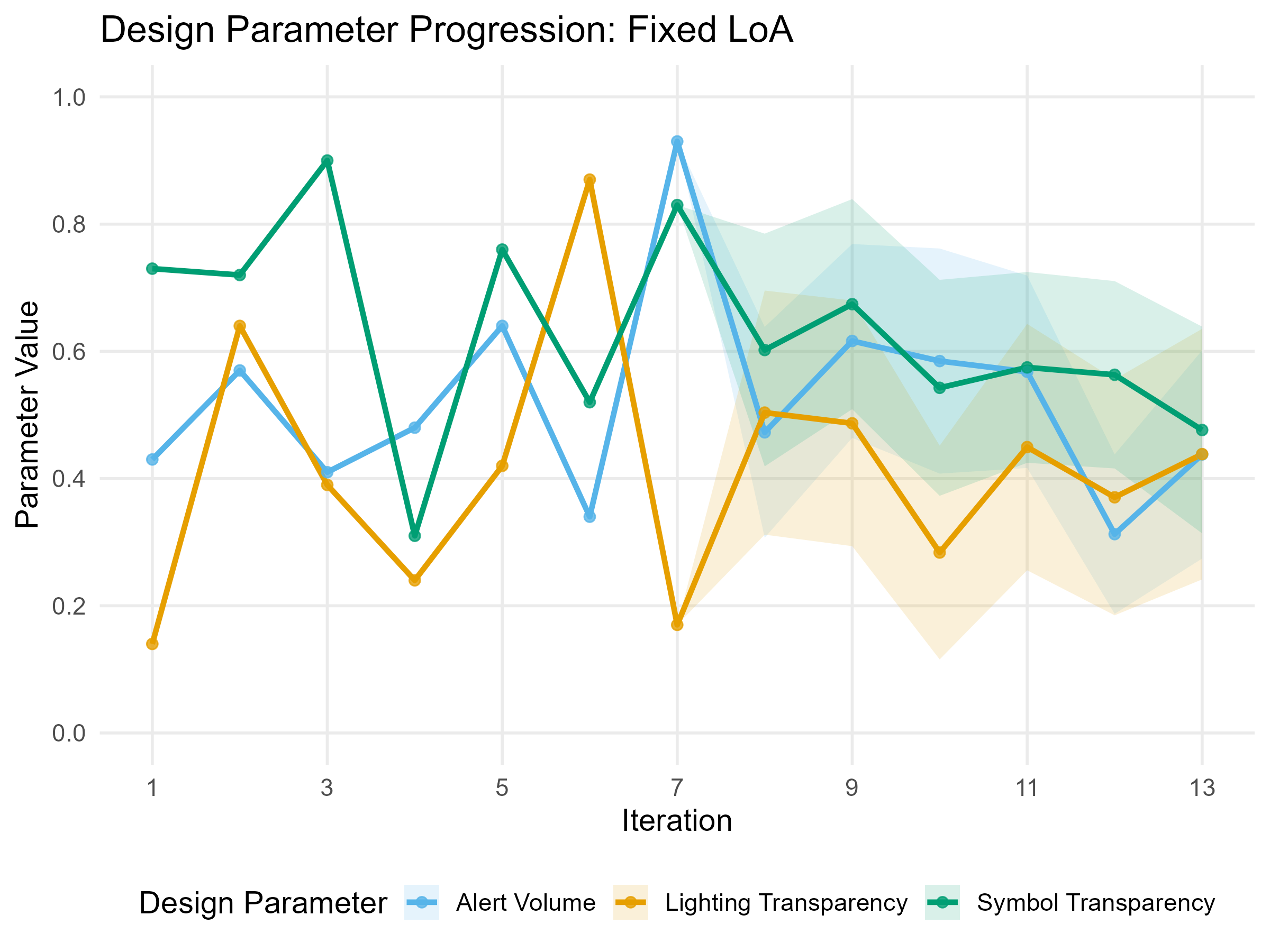}
     \caption{Mean design parameter progression (Fixed Condition)}
     \Description{Mean design parameter progression (Fixed Condition).}
     \label{fig:parameterProgressionFixed}
 \end{figure}

 \begin{figure}
     \centering
     \includegraphics[width=0.5\textwidth]{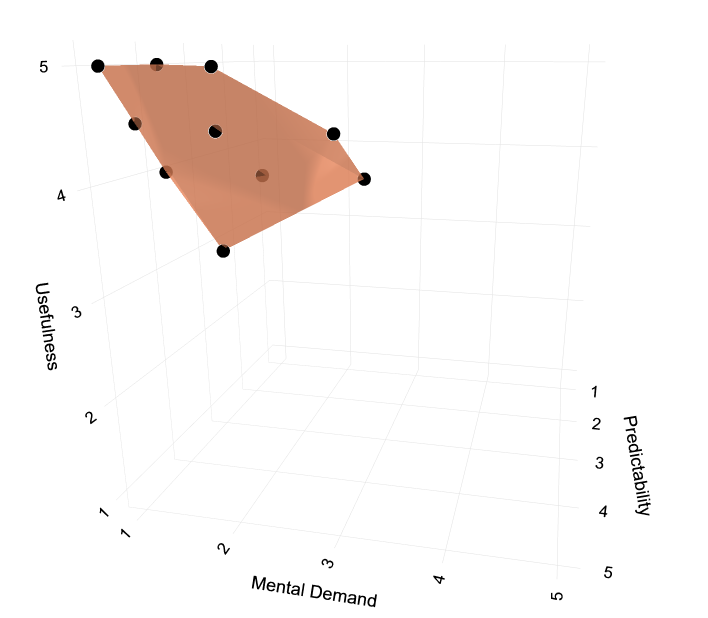}
     \caption{Pareto Front Distribution (Fixed Condition, Side)}
     \Description{Pareto Front Distribution (Fixed Condition, Side).}
     \label{fig:objectivesFixedSide}
 \end{figure}

 \begin{figure}
     \centering
     \includegraphics[width=0.5\textwidth]{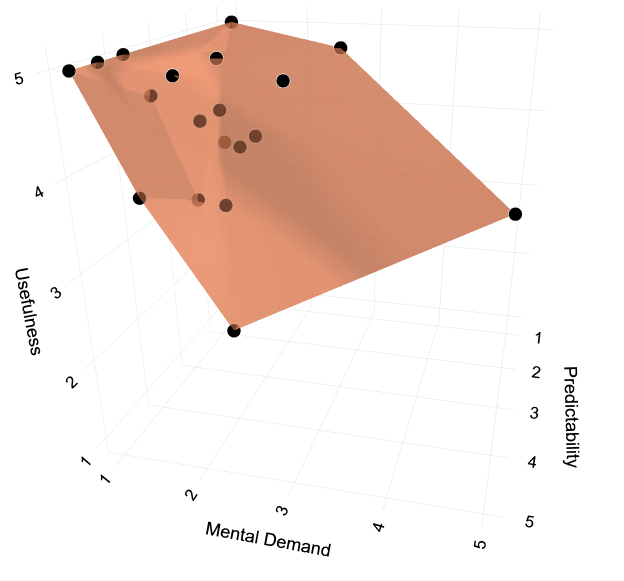}
     \caption{Pareto Front Distribution (Trained Condition, Side)}
     \Description{Pareto Front Distribution (Trained Condition, Side).}
     \label{fig:objectivesTrainedSide}
 \end{figure}

\end{document}